\documentclass[12pt]{article}
\usepackage{amssymb,amsmath}
\usepackage{epsfig}
\usepackage{epstopdf}
\usepackage{latexsym}
\usepackage{graphicx}
\usepackage{subfigure}
\usepackage{booktabs}
\usepackage{bbm}
\usepackage[margin=20pt,small]{caption}

\usepackage[toc]{appendix}

\usepackage{color}
\usepackage{datetime}
\usepackage[
      colorlinks=false,
      linkcolor=darkblue,  
      urlcolor=blue,    
      filecolor=blue,     
      citecolor=red,
linktocpage=true,
      pdfstartview=FitV,
      bookmarksopen=true    
      ]{hyperref}

\ifpdf
\fi

\def\oneone{\rlap 1\mkern4mu{\rm l}}
\def\coeff#1#2{\relax{\textstyle {#1 \over #2}}\displaystyle}
\def\ds{\displaystyle}

\def\ZZ{\mathbb{Z}}

\def\cN{{\cal N}}
\def\cO{{\cal O}}
\def\cP{{\cal P}}

\def\cR{{\cal R}}

\def\eql{=}

\definecolor{cardinal}{rgb}{0.6,0,0}
\definecolor{darkgreen}{rgb}{0,0.5,0}
\definecolor{golden}{rgb}{0.92, 0.7, 0}
\definecolor{midnight}{rgb}{0, 0, 0.5}
\definecolor{darkblue}{rgb}{0.2, 0, 0.8}

 \newcommand{\nc}{\newcommand}
\nc{\bea}{\begin{eqnarray}}
\nc{\eea}{\end{eqnarray}}
\nc{\be}{\begin{equation}}
\nc{\ee}{\end{equation}}
 \nc{\tphi}{\tilde{\phi}}
\nc{\non}{\nonumber}

\begin{document}  
%%%%%%%%%%%%%%%%%%%%%%%%%%%%%%

\begin{titlepage}
 
 \begin{flushright}
UTTG-10-11
\end{flushright}

\medskip
\begin{center} 
{\Large \bf  Minimal Holographic Superconductors from Maximal Supergravity}

\bigskip
\bigskip

{\bf Nikolay Bobev,${}^{(1)}$  Arnab Kundu,${}^{(2)}$
 Krzysztof Pilch,${}^{(3)}$  and Nicholas P. Warner${}^{(3)}$ \\ }
\bigskip
${}^{(1)}$
Simons Center for Geometry and Physics\\
Stony Brook University\\
Stony Brook, NY 11794-3840, USA\\
\vskip 5mm
${}^{(2)}$ 
Theory Group\\ 
Department of Physics\\ 
University of Texas at Austin\\ 
Austin, TX 78712, USA.\\ 
\vskip 5mm
${}^{(3)}$ Department of Physics and Astronomy \\
University of Southern California \\
Los Angeles, CA 90089, USA  \\
\bigskip
nbobev@scgp.stonybrook.edu,~arnab@physics.utexas.edu, \\ 
pilch@usc.edu,~warner@usc.edu  \\
\end{center}

\begin{abstract}

\noindent  
\end{abstract}

We study a truncation of four-dimensional maximal gauged supergravity that provides a realization of the minimal model of a holographic superconductor. We find various flow solutions in this truncation at zero and finite temperature with a non-trivial profile for the charged scalar. Below a critical temperature we find holographic superconductor solutions that represent the thermodynamically preferred phase. Depending on the choice of boundary conditions, the superconducting phase transition is either first or second order. For vanishing temperature we find a flow with a condensing charged scalar that interpolates between two perturbatively stable ${\rm AdS}_4$ vacua and is the zero-temperature ground state of the holographic superconductor. 

\end{titlepage}

%%%%%%%%%%%%%%%%%%%%%%%%%%%%%%%%%%%%%

\tableofcontents

%%%%%%%%%%%%%%%%%%%%%%%%%%%%%%%%%%%%%
\section{Introduction}
%%%%%%%%%%%%%%%%%%%%%%%%%%%%%%%%%%%%%
 
There is now a substantial body of research that uses gauge/gravity duality to describe the physics of strongly coupled field theories with a view towards possible connections with condensed matter and many body physics (see, for example,  \cite{Hartnoll:2009sz,Herzog:2009xv,Horowitz:2010gk}). Most of these constructions have been phenomenological, or ``bottom-up,"  in which the gravity dual of an interesting condensed matter system is postulated {\it ab initio} without using the well-established, but more complicated, holographic dualities that one can derive from open/closed duality in string theory.  While it is certainly interesting to explore the possible physics one can realize in such phenomenological gravitational models, the drawback is that one has limited information about the dual field theory and the completeness and accuracy of the holographic dictionary. On the other hand, if one is able to realize a given phenomenological model as a truncation of ten- or eleven-dimensional supergravity one will have a much better control over the dual field theory because of the well-established holographic dualities. 

There have been attempts to embed the holographic superconductor model\footnote{This model involves a spontaneously broken gauge field in the bulk and therefore is more properly described as a superfluid in the dual field theory. We will adopt the name holographic superconductor since this usage is, by now, standard in the literature.} of \cite{Gubser:2008px, Hartnoll:2008vx,Hartnoll:2008kx} in IIB and eleven-dimensional supergravity  \cite{Denef:2009tp,Gubser:2009qm,Gauntlett:2009dn,Gubser:2009gp,Gauntlett:2009bh,Donos:2011ut,Aprile:2011uq}. These embeddings typically involve studying some particular truncation of the higher-dimensional supergravity on a compact manifold or a consistent truncation of a lower-dimensional gauged supergravity. Since the holographic superconductors are typically not supersymmetric there might be unstable light modes that lie outside the truncation of interest and destabilize the holographic superconductor. This possibility was emphasized in \cite{Bobev:2010ib}, where it was demonstrated that for some of the consistent truncations studied in \cite{Gubser:2009qm,Gauntlett:2009dn,Gubser:2009gp,Gauntlett:2009bh}, which realize the minimal model of a holographic superconductor, there are indeed such unstable modes in the lower-dimensional gauged supergravity. 

Our goal here is to show that one can embed the minimal model of a holographic superconductor, consisting of the metric, a charged scalar with a non-trivial potential and an Abelian gauge field,\footnote{This is sometimes called the Abelian Higgs model.} in the truncation of four-dimensional maximal gauged supergravity studied in \cite{Fischbacher:2010ec}. This truncation of gauged supergravity has an  $SO(3)\times SO(3)$ invariance and contains two ${\rm AdS}_4$ critical points with different cosmological constants. The UV critical point is the $SO(8)$, maximally supersymmetric point that  uplifts to the ${\rm AdS}_4\times S^7$ solution in eleven dimensions. The IR critical point was originally found in \cite{Warner:1983du, Warner:1983vz} and has $SO(3)\times SO(3)$ global symmetry and no supersymmetry. An important fact about the $SO(3)\times SO(3)$ ${\rm AdS}_4$ vacuum is that it is perturbatively stable in the full four-dimensional $\mathcal{N}=8$ gauged supergravity \cite{Fischbacher:2010ec}.\footnote{An uplift of this point to eleven dimensions will be discussed in \cite{to appear}.} 
It should also be emphasized that the $SO(3)\times SO(3)$ point has the lowest value of the cosmological constant of all known stable critical points in four-dimensional gauged supergravity \cite{Fischbacher:2009cj,Fischbacher:2011jx,Fischbacher:2010ec} and the  cosmological constant  is also lower than that of several of the unstable critical points. Therefore the  $SO(3)\times SO(3)$ point has a chance of being the IR attractive critical point for a lot of flows in the theory on the world-volume of M2 branes.

Using the usual Ansatz for a holographic superconductor solution employed in \cite{Hartnoll:2008kx}, we numerically solve the equations of motion in the $SO(3)\times SO(3)$ invariant truncation of gauged supergravity. Depending on the choice of boundary conditions, we find two types of solutions with non-trivial gauge fields and scalar condensates below some critical value of the temperature. These solutions are thermodynamically preferred over the AdS-Reissner-Nordstr\o m (AdS-RN) solution. The phase transition at the critical temperature for one choice of  boundary conditions is second order and the phase diagram looks much like the one studied in \cite{Hartnoll:2008kx,Gauntlett:2009dn,Gauntlett:2009bh}. The phase transition for the other choice of boundary conditions is, however, first order. This is to be contrasted with all other embeddings of holographic superconductors in supergravity for which a second order phase transitions for the condensate was found \cite{Denef:2009tp,Gubser:2009qm,Gauntlett:2009dn,Gubser:2009gp,Gauntlett:2009bh,Donos:2011ut,Aprile:2011uq}. This fact suggests that there is probably no universal behavior of holographic superconductors embedded in higher dimensional supergravity.\footnote{Note that a first order superconducting phase transition was obtained in some of the phenomenological models studied in \cite{Franco:2009yz,Franco:2009if}. To the best of our knowledge these examples have not been embedded in supergravity/string theory.}

To  elucidate the properties of the first order phase transition, we study a family of phenomenological potentials that interpolate between the one in the $SU(4)$ sector studied in \cite{Gauntlett:2009dn,Gauntlett:2009bh}, and the one in the $SO(3)\times SO(3)$ sector.  We explicitly show how the order of the phase transition changes from second to first. While the interpolating potential is ``phenomenological,'' it is important to underline the fact that the end points of this interpolation give potentials that live within fully consistent truncations of eleven-dimensional supergravity and so have well-established holographic interpretations.  This explicitly demonstrates that the physics of the holographic superconductor depends crucially on the truncation of supergravity and its corresponding potential. We also study the zero temperature limit of the solutions and show that there is an emergent conformal symmetry in the IR realized by a domain wall solution interpolating between the two ${\rm AdS}_4$ vacua of the supergravity truncation at hand. Since the IR ${\rm AdS}_4$ vacuum is perturbatively stable,  we have an embedding of the minimal holographic superconductor in gauged supergravity with a stable zero-temperature ground state.
 
It is relatively easy to relate the flows considered here to the Chern-Simons theory on the M2 branes \cite{Aharony:2008ug}.    The $SO(3)\times SO(3)$ is embedded diagonally into the $SO(6)$ $\cR$-symmetry of the ABJM theory so that the six manifest supersymmetries of the ABJM theory decompose as $(3,1) \oplus (1,3)$.  Thus none of the six supersymmetries of the ABJM theory survive in the $SO(3)\times SO(3)$ invariant truncation that we will consider here.  The residual $U(1)$ gauge field, dual to the chemical potential on the M2 brane in our model, is  the $SO(2)$ that commutes with $SO(6)$ inside $SO(8)$ and is thus the $U(1)_b$ baryon number symmetry that is used to make the $\ZZ_k$ orbifold and determines the level of the dual Chern-Simons theory.   The $SO(3)\times SO(3)$ invariant flow of interest involves supergravity scalars that are charged under this $U(1)_b$ and,  as explained in  \cite{Aharony:2008ug}, correspond to 't Hooft, or monopole,  operators.   Thus our flows involve condensates of such monopoles within the dual Chern Simons theory.

In Section 2, we present the action of the supergravity truncation of interest, the Ansatz for the holographic superconductor solutions and the corresponding equations of motion. In Section~3, we study holographic superconductor solutions at zero and finite temperature, present the phase diagrams for the two possible condensates and show that the superconductor solutions are thermodynamically preferred over the AdS-RN solution. Section 4 is devoted to the study of a one-parameter family of phenomenological potentials that interpolate between the potential in the $SU(4)$ sector of gauged supergravity studied in \cite{Gauntlett:2009dn,Gauntlett:2009bh,Bobev:2010ib} and the $SO(3)\times SO(3)$ potential that is the primary focus of this paper. In particular, we show that the order of the phase transition for one of the condensates changes from first to second as we vary the  parameter in the interpolating potential. In Section 5 we discuss the holographic dictionary for our model in some detail and point out that our flows do not realize spontaneous symmetry breaking in the M2 brane field theory but nevertheless they realize holographic superconductors. We conclude in Section 6 with a discussion and possible avenues for further study.

%%%%%%%%%%%%%%%%%%%%%%%%%
\section{The gauged supergravity truncation}
%%%%%%%%%%%%%%%%%%%%%%%%%

%%%%%%%%%%%%%%%%
\subsection{The truncated action}
\label{truncation}
%%%%%%%%%%%%%%%%

We will study a truncation of four-dimensional $\cN=8$  gauged supergravity, \cite{de Wit:1982ig}, to the $SO(3)\times SO(3)$ invariant sector. Under this group, the eight supersymmetries decompose as $(3,1) \oplus (1,3) \oplus (1,1) \oplus  (1,1)$.  In particular, the invariant subsector under one of the $SO(3)$'s is simply the $\cN=5$ gauged supergravity discussed in  \cite{deWit:1981yv}.  We may therefore obtain the theory of interest as the SO(3) invariant sector of  $\cN=5$ gauged supergravity.  The relevant truncation is also discussed in \cite{Fischbacher:2010ec, to appear}. The  theory is $\cN=2$ supergravity coupled to a hypermultiplet and the bosonic sector of the theory consists of the graviton, the graviphoton and  two complex scalar fields, $\zeta_1,\zeta_2$, with charges $\pm 1$ under the $SO(2)$ $\cR$-symmetry:
\begin{equation}\label{gaugteran}
\zeta_1~\longrightarrow~ e^{i\alpha}\zeta_1\,,\qquad 
\zeta_2~\longrightarrow~ e^{-i\alpha}\zeta_2\,.
\end{equation}
There are five complex scalars in  $\cN=5$ gauged supergravity which parametrize the coset $SU(5,1)/U(5)$, and thus the scalars $\zeta_1,\zeta_2$ will be the $SO(3)$-invariant subsector of this and will parametrize the coset 
\begin{equation}\label{thecoset}
{SU(2,1)\over SU(2)\times U(1)}\,.
\end{equation}
With our choice of gauge transformation,  (\ref{gaugteran}), the graviphoton, $A$, gauges the diagonal $U(1)$ subgroup of the denominator $SU(2)$ and the  covariant derivatives of the complex scalars are:
\begin{equation}\label{covder}
\nabla_\mu\zeta_1   ~=~ \partial_{\mu}\zeta_1+ig \,A_\mu\,\zeta_1\,,\qquad 
\nabla_\mu\zeta_2  ~=~ \partial_{\mu}\zeta_2-ig \,A_\mu\,\zeta_2\,,
\end{equation}
where $g$ is the coupling constant of the gauged supergravity. 

Then the truncated bosonic action is\footnote{Note that the coefficient of the $F\wedge F$ term vanishes identically in our truncation.}:
\begin{equation}\label{action}
e^{-1}{\cal L}\eql {1\over 2}R-{1\over 4}F_{\mu\nu}F^{\mu\nu} -g_{\zeta_i\bar\zeta_j}\nabla_\mu\zeta_i\nabla^\mu\bar\zeta_j-\cP\,,
\end{equation}
where the metric on the coset \eqref{thecoset} is given by: 
\begin{equation}\label{quatermetr}
 g_{\zeta_i\bar\zeta_j}d\zeta_i d\bar\zeta_j\eql 
 {d\zeta _1 d\overline\zeta {}_1+d\zeta _2d\overline\zeta {}_2\over 1-|\zeta _1|^2-|\zeta _2|^2} +
{(\zeta _1d\overline\zeta {}_1+\zeta _2d\overline \zeta {}_2)(\overline \zeta {}_1d\zeta _1+\overline \zeta {}_2d\zeta _2)\over (1-|\zeta _1|^2-|\zeta _2|^2)^2}\,,
\end{equation}
and the potential is: 
\begin{equation}\label{potential}
\cP\eql -\coeff{1}{2}\, g^2\, {12 -16(|\zeta_1|^2+|\zeta_2|^2)+3(|\zeta_1|^4+|\zeta_2|^4)+10|\zeta_1|^2|\zeta_2|^2\over  (1-|\zeta_1|^2-|\zeta_2|^2)^2}\,.
\end{equation}

Since we are going to need the holographic dictionary of the $\cN=8$ theory, it is important to relate the scalars $\zeta_1$ and   $\zeta_2$ to those of the $\cN=5$ theory and thereby to those the $\cN=8$ theory.  Define:
\begin{equation}\label{otherparam}
\phi_1 ~\equiv~ \coeff{1}{\sqrt{2}}\, (\zeta_1  + \zeta_2)  \,, \qquad   \phi_2 ~\equiv~ \coeff{i}{\sqrt{2}}\, (\zeta_1  - \zeta_2)  \,,
\end{equation}
then these scalars have gauge transformations as a two-dimensional vector of $SO(2)$ and may be viewed as two of the five complex scalars of the $\cN=5$ theory.  One can easily verify that with this change of variables the scalar potential, (\ref{potential}), is precisely the $SO(3)$ invariant truncation of the potential of the $\cN=5$ theory \cite{deWit:1981yv}. The real parts of the $\phi_i$ are scalars and the imaginary parts are pseudoscalars. Thus the former are  dual to boson bilinears in the M2 brane theory and the latter are dual to fermion bilinears.  

In addition to the  maximally supersymmetric $SO(8)$ critical point for which $\zeta_1=\zeta_2=0$, there are 
the non-trivial $SO(3)\times SO(3)$-invariant critical points  at:
\begin{equation}\label{ctpt1}
\zeta_i\eql0\,,\qquad \zeta_{j}\eql\pm {2\over\sqrt5}\,,\quad i\not=j=1,2\,,
\end{equation}
which correspond to
\begin{equation}\label{ctpt2}
\phi_1~=~  i \, \epsilon_ 1 \, \phi_2 ~=~ \epsilon_ 2 \,   \sqrt {2\over5}\,, 
\end{equation}
with  $\epsilon_ 1^2=\epsilon_ 2^2 = 1$.  As shown in \cite{Fischbacher:2010ec}, this critical point is not supersymmetric but is still perturbatively stable in the full $\mathcal{N}=8$ gauged supergravity, that is, all seventy scalars have masses above the BF bound \cite{Breitenlohner:1982jf}. 

The potential, (\ref{potential}), is invariant under $\zeta_1\leftrightarrow \pm\zeta_2$ and under $\zeta_i \to - \zeta_i$ for $i=1,2$ separately.  Indeed, one can show that it is consistent with all the equations of motion derived from \eqref{action} to set $\zeta_1=0$ and we will do so henceforth. Note that  setting $\zeta_1 =0$ sets $\phi_2 = - i \phi_1$ and thus locks together scalars and pseudoscalars.  Since the scalars and pseudoscalars lie in different $SO(8)$ representations of the $\cN=8$   supergravity theory, setting $\zeta_1=0$ cannot be induced as a part of the gauge symmetry and thus the $\zeta_1 \to -\zeta_1$ symmetry that allows this identification should be viewed as an ``accidental symmetry'' of the action.  This symmetry does, however, make the analysis of the flow to the non-trivial critical point far simpler.

 It is convenient to perform the following change of variables:
\begin{equation}\label{coorrchange}
\zeta_2\eql \tanh \lambda\,e^{i\varphi}\,,
\end{equation}
and this simplifies the action to the form that we will use throughout the rest of the paper:
\begin{equation}\label{actionsimple}
e^{-1}{\cal L}\eql {1\over 2}R-{1\over 4}F_{\mu\nu}F^{\mu\nu} - \partial_{\mu}\lambda\partial^{\mu}\lambda - \ds\frac{\sinh^2(2\lambda)}{4}(\partial_{\mu}\varphi -g A_{\mu})(\partial^{\mu}\varphi -g A^{\mu}) - \mathcal{P}\,,
\end{equation}
with the  potential:
\begin{equation}\label{SO3pot}
\mathcal{P} = - g^2 \,\big(6\, \cosh^4(\lambda) ~+~ 6\, \cosh^2(\lambda) \, \sinh^2(\lambda)~+~\coeff{3}{2} \, \sinh^4(\lambda)\big) \,.
\end{equation}
The critical points of the potential are at
\begin{equation}\label{}
\lambda = 0~, \qquad\qquad \lambda\eql \log(2+\sqrt{5})\,,
\end{equation}
having $SO(8)$ and $SO(3)\times SO(3)$ global symmetry respectively.

 %%%%%%%%%%%%%%%%
\subsection{The equations of motion}
%%%%%%%%%%%%%%%%

The action in (\ref{actionsimple}) is that of a charged scalar with a non-trivial potential coupled to gravity. These are the minimal ingredients of the holographic superconductor model studied in \cite{Gubser:2008px,Hartnoll:2008vx,Hartnoll:2008kx}. Below we will show that indeed this consistent truncation of four-dimensional gauged supergravity admits solutions that can be interpreted as holographic superconductors. To do this we take the following Ansatz for the metric, the gauge field and the scalar\footnote{For ease of comparison, we have used the same Ansatz as in \cite{Hartnoll:2008kx,Gauntlett:2009dn,Gauntlett:2009bh}.}:
\begin{equation}
ds^2 = -G(r) e^{-\chi(r)}dt^2 + r^2(dx_1^2+dx_2^2) +\ds\frac{dr^2}{G(r)}~, \qquad A = \Psi(r) dt~, \qquad \lambda=\lambda(r)~.
\end{equation} 
It is straightforward to substitute this Ansatz in the equations of motion derived from \eqref{actionsimple} and find a system of ordinary differential equations that govern  radial flows (that is, flows that  depend only upon $r$). The $(t,r)$ component of Einstein equations leads to the following equation: 
\begin{equation}
\sinh^2(2\lambda)\Psi  \frac{d\varphi}{dr}  ~=~ 0~.
\end{equation} 
Since we are interested in solutions with non-trivial profiles for $\lambda$ and $\Psi$ we choose to solve this by taking $\varphi$ to be a constant and, because of the symmetry, we can take this to be zero.   With this choice, the rest of the equations of motion reduce to:
\begin{eqnarray}
&& \chi'+2 r (\lambda')^2 + \ds\frac{g^2 r e^{\chi} \sinh^2(2\lambda)\Psi^2}{2G^2} = 0~, \label{Ein1}\\
&& (\lambda')^2 +\ds\frac{G'}{rG}+ \ds\frac{e^{\chi}(\Psi')^2}{2G} + \ds\frac{\mathcal{P}}{G}  + \ds\frac{1}{r^2} + \ds\frac{g^2 e^{\chi} \sinh^2(2\lambda)\Psi^2}{4G^2} = 0~,\label{Ein2}\\
&& \Psi'' + \left(\ds\frac{2}{r}+\ds\frac{\chi'}{2}\right)\Psi' - \ds\frac{g^2 \sinh^2(2\lambda)\Psi}{2G} = 0~, \label{Max}\\
&&\lambda'' + \left(\ds\frac{2}{r}-\ds\frac{\chi'}{2} + \ds\frac{G'}{G}\right)\lambda' -  \ds\frac{1}{2G} \ds\frac{d\mathcal{P}}{d\lambda}  + \ds\frac{g^2 e^{\chi} \sinh(4\lambda)\Psi^2}{4G^2}  = 0~, \label{KG}
\end{eqnarray}
where  $'$ denotes $d/dr$.
Equations \eqref{Ein1} and \eqref{Ein2} are appropriate linear combinations of the $tt$ and $rr$ components of Einstein equations. Equation \eqref{Max} is the $t$ component of Maxwell equations and \eqref{KG} is the equation of motion for the scalar $\lambda$. The $x_ix_i$ components of the Einstein equations lead to equations that can be derived from  \eqref{Ein1} and \eqref{Ein2} and are therefore not independent. We will numerically solve  equations (\ref{Ein1})--(\ref{KG})   in Section \ref{Section3}.  

Since we will be looking for black-hole solutions in a static metric, the horizon will be the zero locus of $G(r)$.  Specifically, the horizon is located at $r=r_H$  where $G(r) \sim \cO(r-r_H)$.   Regularity also requires that  $\Psi(r) \sim \cO(r-r_H)$ at the horizon.   The temperature of the solution can be computed in the standard way by imposing regularity of the Euclidean metric near $r=r_H$.   Indeed, if one uses  (\ref{Ein2})  and extracts the simple pole term at $r=r_H$ one obtains a simple expression for the temperature: 
\begin{equation}\label{temp}
T ~=~- \frac{r_H}{8\pi}\, ( 2 \mathcal{P} e^{-\chi/2} ~+~ (\Psi')^2 e^{\chi/2})|_{r=r_H} \,.
\end{equation} 
%

 %%%%%%%%%%%%%%%%%%%%%%%%%%%%%%%%%%%%%%%%%%%%%%%%%%%%%%%%%%%%%%%%%%%%%%%%%%
\section{Holographic superconductors}
\label{Section3}
%%%%%%%%%%%%%%%%%%%%%%%%%%%%%%%%%%%%%%%%%%%%%%%%%%%%%%%%%%%%%%%%%%%%%%%%%%

%%%%%%%%%%%%%%%%%%%%%%%%%%%%%%%%%%% 
\subsection{Solutions at finite temperature}
%%%%%%%%%%%%%%%%%%%%%%%%%%%%%%%%%%%

We will show below that, at finite temperature, there are, in general, two types of solutions to the equations of motion. One is the familiar AdS-RN black hole that exists for all values of the temperature.\footnote{Strictly speaking this is only true when one studies black holes with flat horizons, as we do here. In global coordinates there is a Hawking-Page phase transition at sufficiently low temperature.} Below some critical value of the temperature we  find a new branch of solutions that have scalar hair and are thermodynamically preferred over the AdS-RN solution.

The equations of motion, \eqref{Ein1}--\eqref{KG}, and the background fields have the following scaling symmetries that need to be fixed before constructing a solution:  
\begin{equation}
\label{scaling1}
t \to \beta_1 t~, \qquad\qquad \chi \to \chi + 2\log \beta_1~, \qquad\qquad \Psi \to \beta_1^{-1} \Psi~,
\end{equation} 
\begin{equation}
\label{scaling2}
t \to \beta_2 t~, \qquad r \to \beta_2 r~, \qquad\qquad g\to \beta_2^{-1} g~,
\end{equation} 
\begin{equation} \label{scaling3}
(t,x_1,x_2) \to \beta_3^{-1} (t,x_1,x_2)~,  \qquad r \to \beta_3 r~,  \qquad \Psi \to \beta_3 \Psi~, \qquad G \to \beta_3^2 G~,
\end{equation} 
where $(\beta_1,\beta_2,\beta_3)$ are real scaling parameters. These symmetries can be used to choose arbitrary values for the location of the  horizon, the coupling constant of gauged supergravity, $g$, and the asymptotic value of the metric function $\chi_{\infty}=\ds\lim_{r\to\infty}\chi(r)$. We will chose the following values
\begin{equation}
r_H=1~, \qquad\qquad g=1~, \qquad\qquad \lim_{r\to\infty}\chi = 0~.
\end{equation} 
%
 
%%%%%%%%%%%%%%%%%%%%
\subsubsection{The AdS-RN black hole}
%%%%%%%%%%%%%%%%%%%%

The AdS-RN solution is simply obtained by setting $\lambda = 0$ and $\chi = 0$. The metric function and the gauge fields are given by\footnote{From now on we will fix the coupling constant of gauged supergravity $g=1$. This also sets the scale of the ${\rm AdS}_4$ critical points of the scalar potential.}
\begin{eqnarray}
G = 2r^2 - \frac{1}{r} \left( 2r_H^3 + \frac{\rho^2}{ 2r_H}\right) + \frac{\rho^2}{2 r^2} \ , \qquad \Psi =  \rho \left( \frac{1}{r_H} - \frac{1}{r}\right) \ .
\end{eqnarray}
The chemical potential is given by the potential difference between the horizon and infinity, $\mu = \rho/r_H$, the charge density in the dual field theory is $\rho$ and the temperature of the black hole is given by:
\begin{eqnarray}
T = \frac{12 r_H^4 - \rho^2}{8 \pi r_H^3} \ .
\end{eqnarray}
The extremal AdS-RN black hole has $T=0$ and therefore $\rho = 2 \sqrt{3} r_H^2$. The metric function is then:
\begin{eqnarray}
G =\ds\frac{2}{r^2} (r-r_H)^2 (r^2 +2 r r_H^2+ 3r_H^2) \ .
\end{eqnarray}
It is clear from this expression that the two horizons of the AdS-RN black hole coincide at $T=0$. It is not hard to show that the extremal AdS-RN black hole is a solution interpolating between ${\rm AdS}_4$ (as $r\to \infty$) and ${\rm AdS}_2\times \mathbb{R}^2$ (as $r\to0$).

%%%%%%%%%%%%%%%%%%%%
\subsubsection{Hairy black hole}
%%%%%%%%%%%%%%%%%%%%

We will now look for solutions of the equations of motion  \eqref{Ein1}--\eqref{KG} that have a non-trivial profile for the scalar. We will use a numerical shooting technique and we will impose initial conditions in the IR, that is, at  the black hole horizon, and read off the solution at asymptotic infinity. The series solution near the horizon has the expansion\footnote{We have put $\Psi^0 =0$ as required by regularity.}:
\begin{align}
\begin{split}
& \chi = \chi^0 + \chi^1 (r-r_H)  + \ldots \ , \\[0pt]
& \lambda = \lambda^0 + \lambda^1 (r-r_H)  + \ldots \ , \\[0pt]
& \Psi =  \Psi^1 (r-r_H) + \Psi^2 (r-r_H)^2 + \ldots  \ , \\[0pt]
& G = G^1 (r-r_H) + \ldots \ .
\end{split}
\end{align}
Substituting this into the equations of motion yields four independent algebraic equations relating the seven parameters. Thus we really have three independent parameters which give us the initial conditions at the horizon. We can choose the following to be the independent ones:
\begin{eqnarray}
\chi^0 \ , \quad \lambda^0 \ , \quad \Psi^1 \ .
\end{eqnarray}

Our numerical scheme is as follows: First we fix $g =1$ and $r_H=1$ using two of the scaling symmetries of the equations of motion \eqref{scaling2} and \eqref{scaling3}. We also fix $\chi^0 = 4$ (one could also pick any other value), and this ultimately generates some non-zero value for $\chi_\infty$ which we then shift, via the scaling symmetry (\ref{scaling1}),  to $\chi_\infty=0$. Then we scan the parameter space $\{\lambda^0, \Psi^1\}$ and generally obtain a two-parameter family of solutions. However, we need to fix the asymptotic boundary behaviour of the scalar field $\lambda$. The asymptotic behaviour of $\lambda$ is given by
\begin{equation}
\lambda \sim \ds\frac{\lambda_1}{r} + \ds\frac{\lambda_2}{r^2} + \ldots~,
\label{UVexplambda}
\end{equation}
and we will choose either $\lambda_1 = 0$ or $\lambda_2 = 0$. Making this choice ultimately leaves us with a one-parameter family of solutions which we then choose to parametrize in terms of the temperature.

To calculate the thermodynamic properties of our solutions we will also need the linearized solution in the UV. Near the boundary of ${\rm AdS}_4$, $r\to \infty$, the solution has the following expansion:
\begin{align}
\label{sbdry1}
\begin{split}
& G = G_{-2} r^2 + G_{-1} r + G_0 + \frac{G_1}{r} + \ldots \ , \\[0pt]
& \chi = \chi_{\infty} + \frac{\chi_1}{r} + \frac{\chi_2}{r^2} + \frac{\chi_3}{r^3} + \frac{\chi_4}{r^4} + \ldots \ , \\[0pt]
& \lambda = \frac{\lambda_1}{r} +  \frac{\lambda_2}{r^2} +  \frac{\lambda_3}{r^3}   + \ldots \ , \\[0pt]
& \Psi = \mu - \frac{\rho}{r} + \frac{\Psi_2}{r^2} +  \frac{\Psi_3}{r^3} +  \frac{\Psi_4}{r^4}  + \ldots \ .
\end{split}
\end{align}
The parameters $\{\chi_{\infty}, \lambda_1, \lambda_2, \mu, \rho \}$ contain physical information and, after substituting the series expansion in the equations of motion and solving up to $\mathcal{O} (1/r^6)$, we can solve for the other coefficients 
\begin{align}
\label{sbdry2}
\begin{split} 
& G_{-2} = 2 \ , \quad G_{-1} = 0 \ , \quad G_0 = 2 \lambda_1^2 \ , \quad G_1 = 3 \lambda_1 \lambda_2 - \epsilon \ , \\[0pt]
& \chi_1 = 0 \ , \quad \chi_2 = \lambda_1^2 \ , \quad \chi_3 = \frac{8}{3} \lambda_1 \lambda_2 \ , \quad \chi_4 = \frac{1}{4} \left( \lambda_1^4 + 8 \lambda_2^2 - e^{\chi_{\infty}} \lambda_1^2 \mu^2 \right) \ ,  \\[0pt]
& \lambda_3 = \frac{1}{24} \lambda_1 \left( 2 \lambda_1^2 - 3 e^{\chi_{\infty}} \mu^2 \right) \ , \qquad \Psi_2 =  \frac{\mu}{2}\,  \lambda_1^2 \ , \qquad \Psi_3 =  \frac{\mu}{3}\, \lambda_1 \lambda_2 \ .  
\end{split}
\end{align}
Note that in $G_1$ we have introduced an independent parameter $\epsilon$ which corresponds to the mass of the hairy black hole. This parameter is not determined in terms of $\{\chi_{\infty}, \lambda_1, \lambda_2, \mu, \rho \}$. 

\begin{figure}[!ht]
\begin{center}
\includegraphics[width=8.5cm]{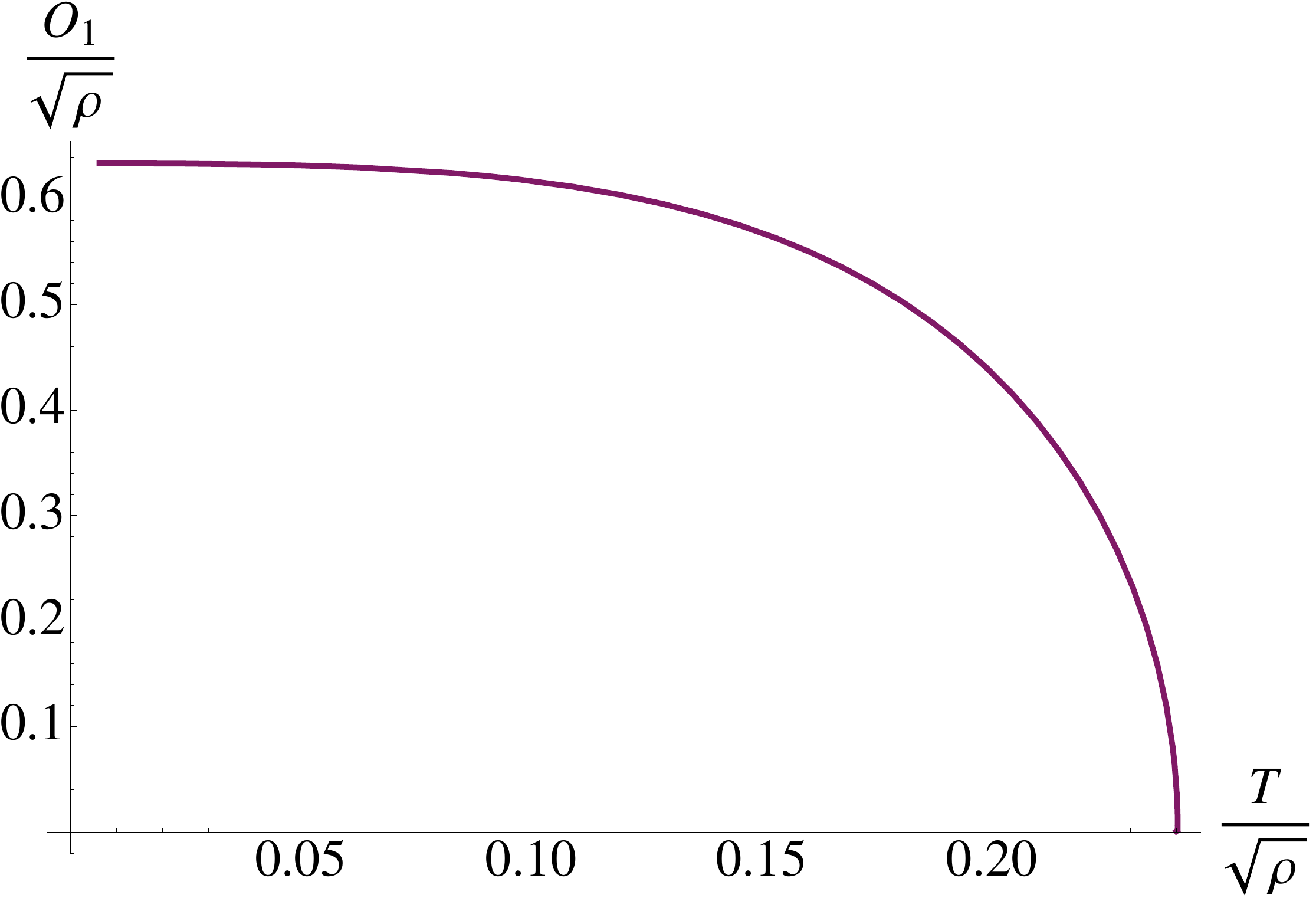}
\caption{ \small The condensate, $O_1\sim \lambda_1$, as a function of temperature with boundary condition $\lambda_2 =0$. The phase transition is second order and happens at $T_c/\sqrt{\rho} \approx 0.2403$.}
\label{O1SO3}
\end{center}
\end{figure}

\begin{figure}[!ht]
\begin{center}
\includegraphics[width=8.5cm]{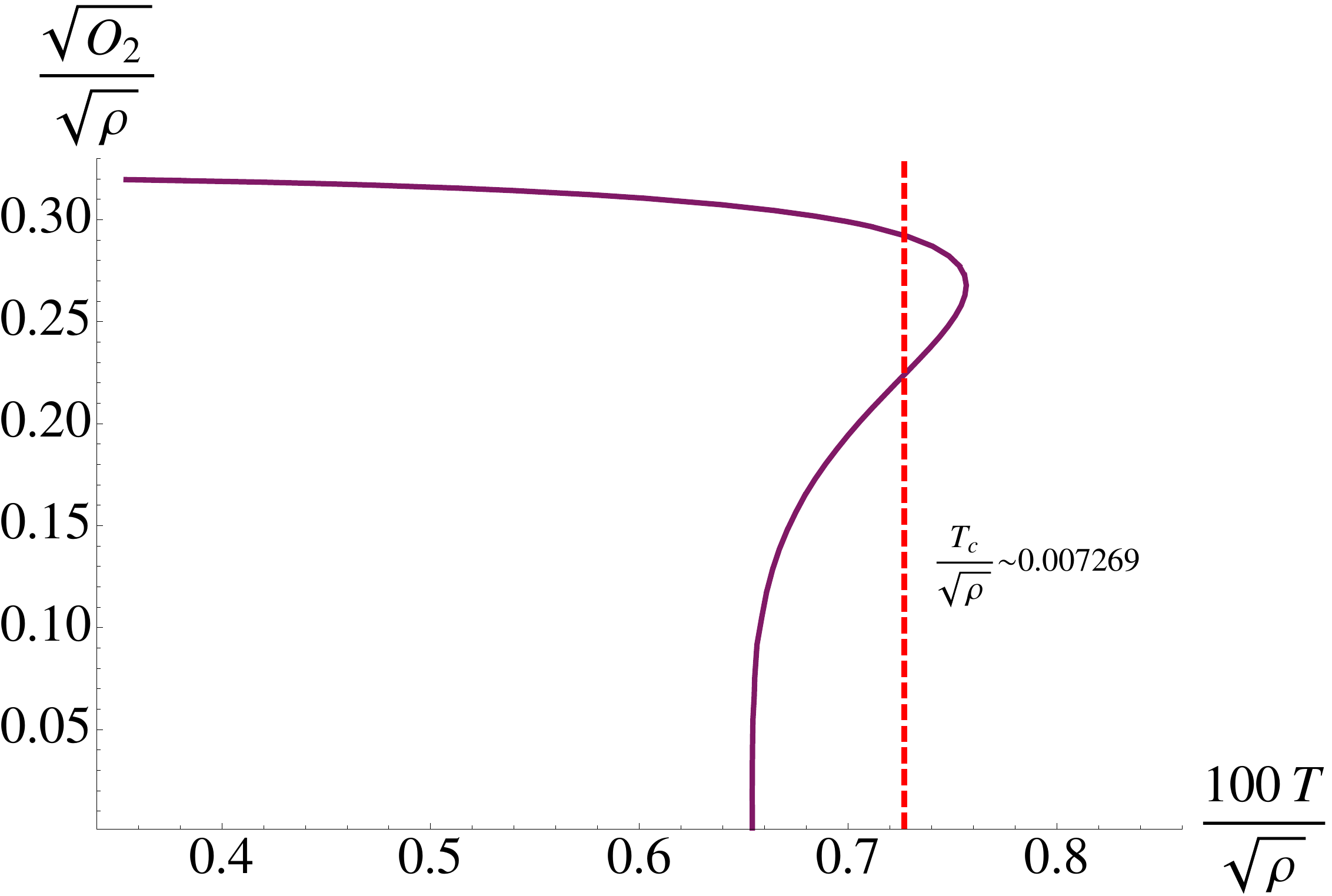}
\caption{ \small The condensate, $O_2\sim \lambda_2$, as a function of temperature with boundary condition $\lambda_1 =0$. The critical temperature at which there is a first order phase transition with a discontinuous jump in the value of the condensate is determined from the free energy computation shown in Fig. \ref{deltaFO2}  and  the value is shown here by the dashed vertical line. }
\label{O2SO3}
\end{center}
\end{figure}

With all these preliminaries, we can now solve for the hairy black hole background. Depending on our choice, we can either have the condensate $O_1 \sim \lambda_1 \not = 0$ or $O_2 \sim \lambda_2 \not = 0$. There exists a one-parameter family of both these types of solutions. This family of solutions generate the corresponding condensate as a function of the temperature. As in \cite{Hartnoll:2008vx, Hartnoll:2008kx} we will work in the fixed charge ensemble otherwise known as the canonical ensemble. The condensates and the temperature are all dimensionful quantities and we will measure them in units of the charge density $\rho$. The dependences are shown in Fig.~\ref{O1SO3} and Fig.~\ref{O2SO3} respectively. The phase transition for $O_1$ is second order and the one for $O_2$ is first order. To determine whether the hairy black hole solutions are thermodynamically preferred one has to compute their free energy and show that it is lower than that of the AdS-RN solution. Computing the  free energy also  enables us to obtain the critical temperature for the first order phase transition for the $O_2$ condensate. This will be the subject of the next section. 
 
%%%%%%%%%%%%%%%%%%%%%%%%%%%%%%%%%%%%%
%%%%%%%%%%%%%%%%%%%%%%%%%%%%%%%%%%%%%

%%%%%%%%%%%%%%%%%%%%%%%%%%%%%%%%%%%%%
%%%%%%%%%%%%%%%%%%%%%%%%%%%%%%%%%%%%%

%%%%%%%%%%%%%%%%%%%%%%%%%%%%%%
\subsubsection{Thermodynamics}
%%%%%%%%%%%%%%%%%%%%%%%%%%%%%%

One can use standard holographic technology to compute the free energy of our solutions. Since this procedure is well known we will omit many of the calculational details here and will just present the relevant final formulae (for more details see for example \cite{Hartnoll:2008kx,to appear}). The basic idea is that the Gibbs free energy in the grand canonical ensemble of the dual field theory is given by the renormalized on-shell Euclidean supergravity action.\footnote{For a review on holographic renormalization see \cite{Skenderis:2002wp}.} The final result depends on the UV boundary condition one imposes, that is, whether one keeps $\lambda_1$ or $\lambda_2$ in \eqref{UVexplambda} fixed.  We find
\begin{align}
\begin{split}
\frac{\Omega}{T} & =  - \frac{1}{2} \int d^3 x e^{-\chi_{\infty}/2} \left( \epsilon - 5 \lambda_1 \lambda_2 \right) \quad {\rm with} \quad   \lambda_1 = \text{fixed} \ , \\
\frac{\Omega}{T} & =  - \frac{1}{2} \int d^3 x e^{-\chi_{\infty}/2} \left( \epsilon - 13 \lambda_1 \lambda_2 \right) \quad {\rm with} \quad  \lambda_2 = \text{fixed} \ ,
\end{split}
\end{align}
where $\Omega$ is the Gibbs free energy in the grand-canonical ensemble, $T$ is the temperature \eqref{temp}, and $\epsilon$ was defined in \eqref{sbdry2}. For all the hairy black hole solutions we study we impose either $\lambda_1 = 0$ or $\lambda_2 = 0$, hence the second term in the final expression for $\Omega$ above does not contribute and we end up with a formula identical to the one obtained in \cite{Hartnoll:2008kx}.

When we constructed the hairy black hole solutions we fixed the charge in the dual field theory and this corresponds to choosing the canonical ensemble. The Helmholtz free energy in the canonical ensemble is given by
\begin{eqnarray} \label{freehairy}
F_{\text{hairy}} = \Omega + \mu \rho \int d^2x = \left( - \frac{\epsilon}{2} +  \mu\rho \right) V_{{\mathbb R}^2} \ , 
\end{eqnarray}
where $V_{{\mathbb R}^2}$ is the spatial volume in the $(x_1,x_2)$ plane and we further note that the physical temperature is given by $\int d\tau e^{-\chi_\infty/2} = 1/T$, where $\tau$ is the Euclidean time. 

\begin{figure}[!ht]
\begin{center}
\includegraphics[width=8.5cm]{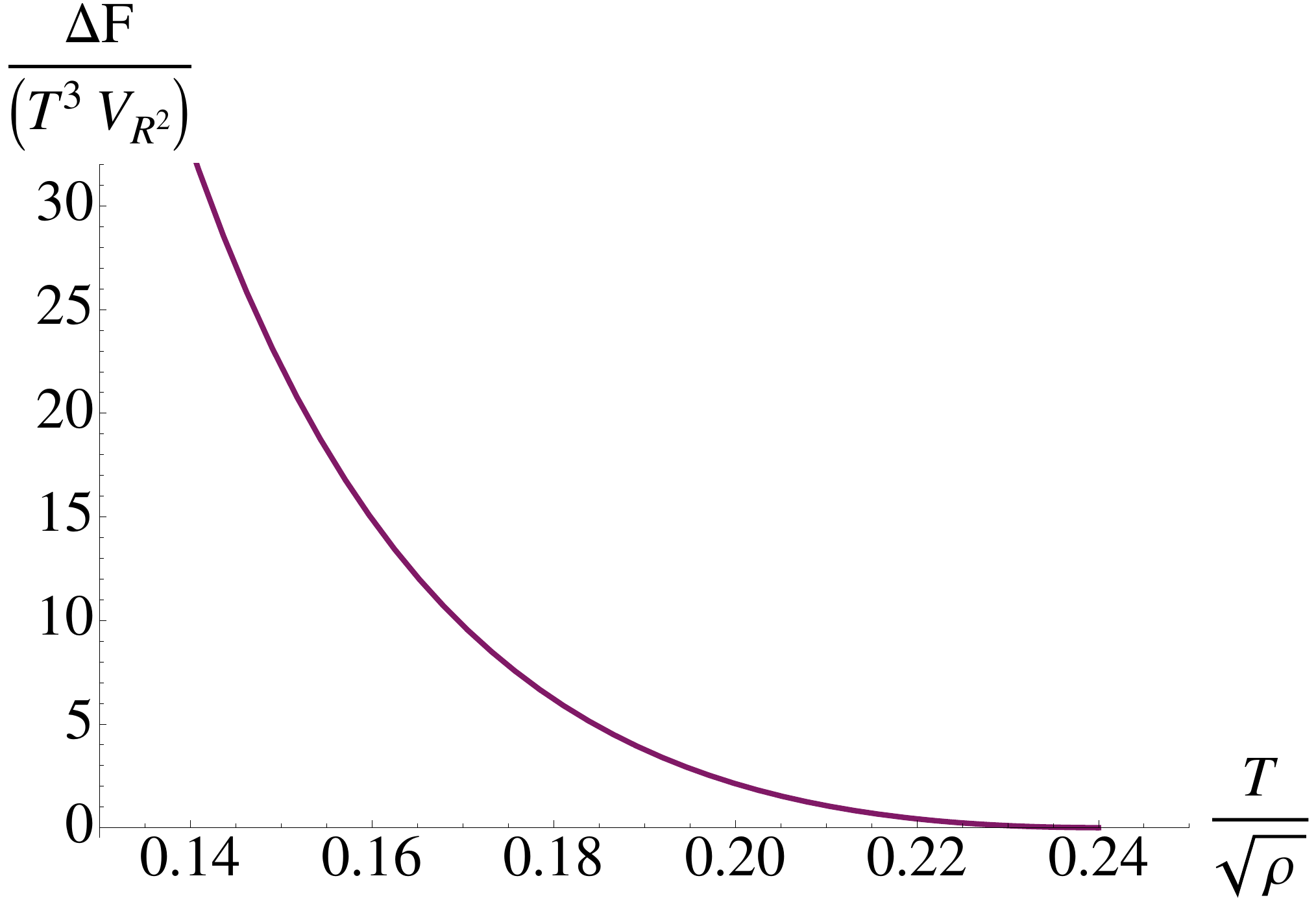}
\caption{ \small The difference in free energy between the AdS-RN and the hairy black hole solutions for the hairy solutions with a non-zero $O_1$ condensate. The critical temperature, $T_c/\sqrt{\rho} \approx 0.2403$, is determined by where $\cO_1$ vanishes (see Fig. \ref{O2SO3}) and the hairy black hole solution has a lower free energy for $T<T_c$ and the phase transition is second order.}
\label{deltaFO1}
\end{center}
\end{figure}

\begin{figure}[!ht]
\begin{center}
\includegraphics[width=8.5cm]{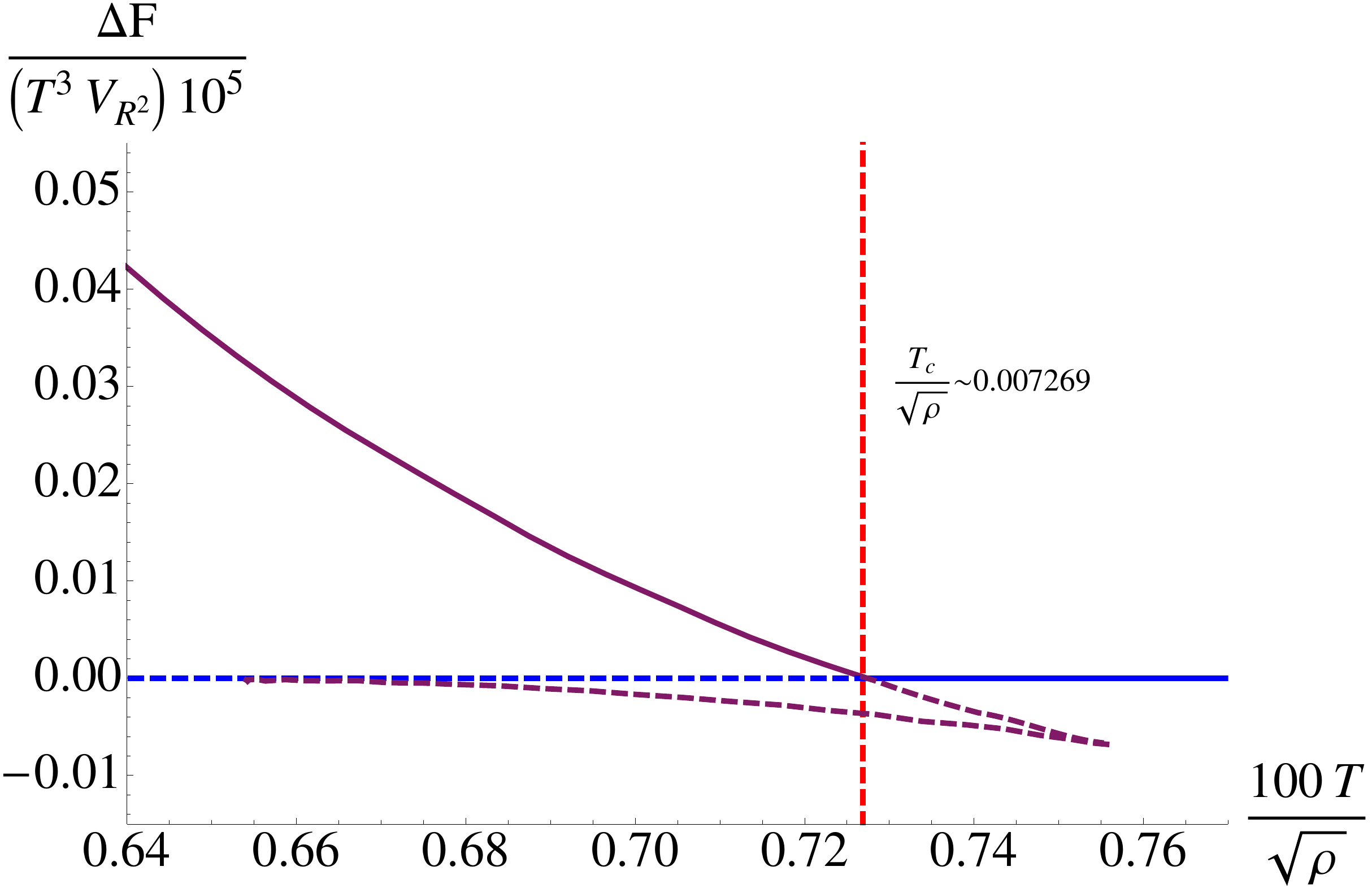}
\caption{ \small The difference in free energy between the AdS-RN and the hairy black hole solutions for the hairy solutions with a non-zero $O_2$ condensate.    The blue (horizontal) curve is the free energy for $T>T_c$ and the kink in the free energy at $T=T_c$ indicates a first order phase transition. The metastable phase is depicted with the dashed maroon (lower) curve.}
\label{deltaFO2}
\end{center}
\end{figure}

As in \cite{Hartnoll:2008kx}, we can also easily compute the Helmholtz free energy in the canonical ensemble for the AdS-RN solution 
\begin{eqnarray}\label{freeRN}
F_{\text{RN}} = \Omega + \mu \rho V_{{\mathbb R}^2} = \frac{1}{r_H} \left( - r_H^4 + \frac{3}{4} \rho^2 \right) V_{{\mathbb R}^2} \ .
\end{eqnarray}
To decide whether the hairy back hole solution is thermodynamically preferred over the AdS-RN black hole we have computed the difference in free energy between the two solutions, $\Delta F= F_{\text{AdS-RN}} - F_{\text{hairy}}$. The thermodynamically preferred branch will have lower free energy. The result for the $O_1$ and $O_2$ condensates are plotted in Fig.~\ref{deltaFO1} and Fig.~\ref{deltaFO2} respectively. It is clear that for $T<T_c$ the hairy black holes have lower free energy and therefore are the thermodynamically preferred phase of the system. The plot for the $O_2$ free energy also clearly demonstrates that there is a ``kink" in the free energy, \textit{i.e.} a discontinuity in the first derivative, at $T=T_c$ and therefore the corresponding phase transition is first order.

%%%%%%%%%%%%%%%%%%%%%%%%%%%%%%%%%%%%%
%%%%%%%%%%%%%%%%%%%%%%%%%%%%%%%%%%%%%

%%%%%%%%%%%%%%%%%%%%%%%%%%%%%%%%%%%%%
%%%%%%%%%%%%%%%%%%%%%%%%%%%%%%%%%%%%%

%%%%%%%%%%%%%%%%%%%%%%%%%%%%%%%%%%%%%%%%%%%%%%%%%%%%%%%%%%%%%%%%%%%%%%%%%%
\subsection{Zero-temperature solutions}
%%%%%%%%%%%%%%%%%%%%%%%%%%%%%%%%%%%%%%%%%%%%%%%%%%%%%%%%%%%%%%%%%%%%%%%%%%

\begin{figure}[t]
 \centering
    \includegraphics[width=6.5cm]{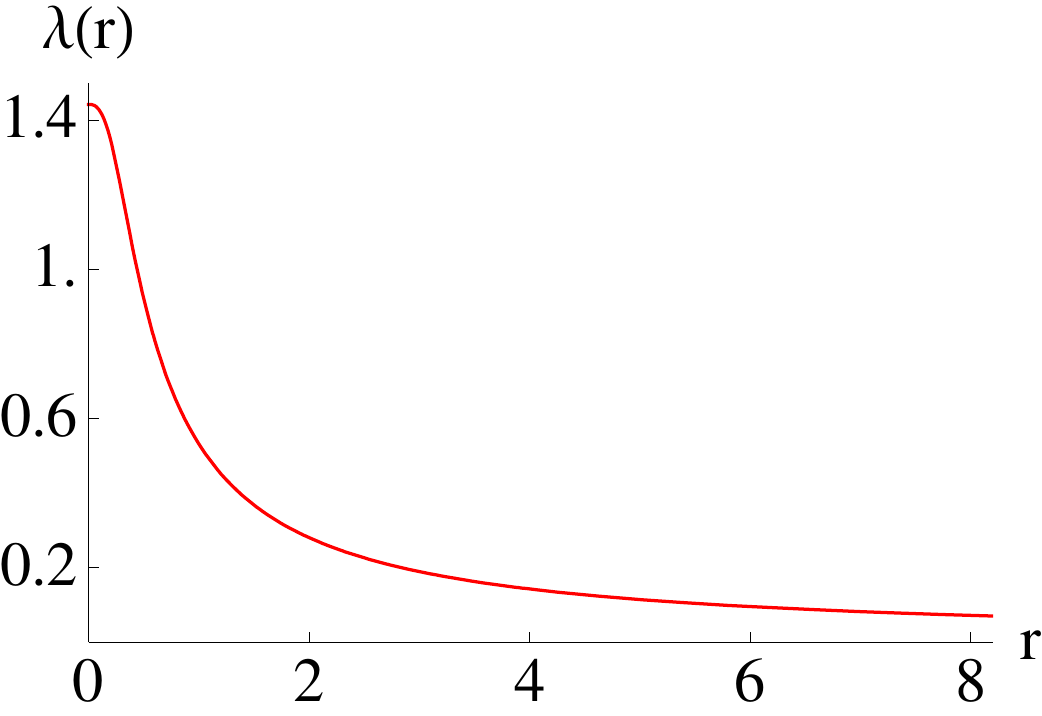}\qquad\qquad
    \includegraphics[width=6.5cm]{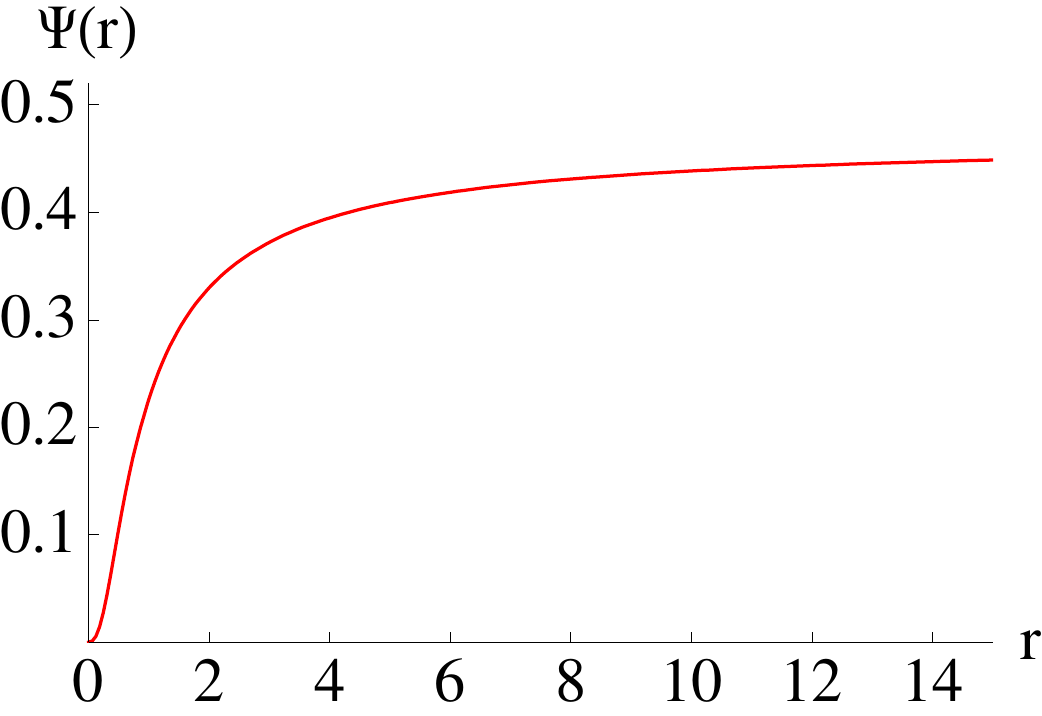}
    \caption{ \small The scalar, $\lambda$, and the gauge field, $\Psi$, as functions of $r$ for the $T=0$ flow with $\lambda_2=0$ boundary condition.}
\label{O1T=0}
\end{figure}

\begin{figure}[t]
 \centering
    \includegraphics[width=6.5cm]{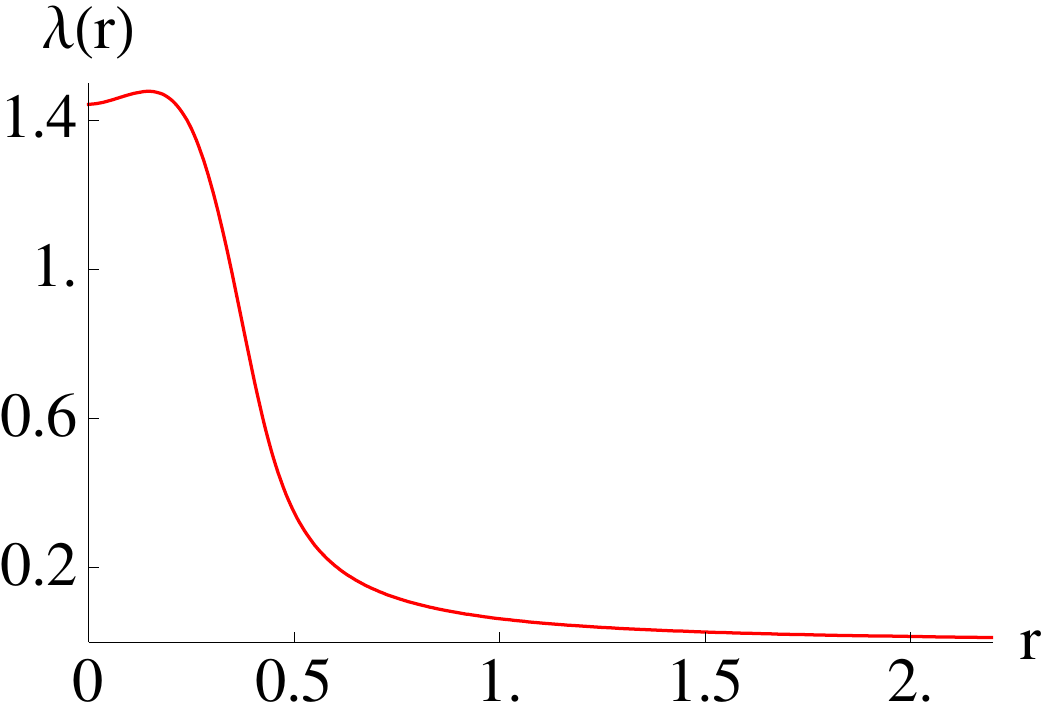}\qquad\qquad
    \includegraphics[width=6.5cm]{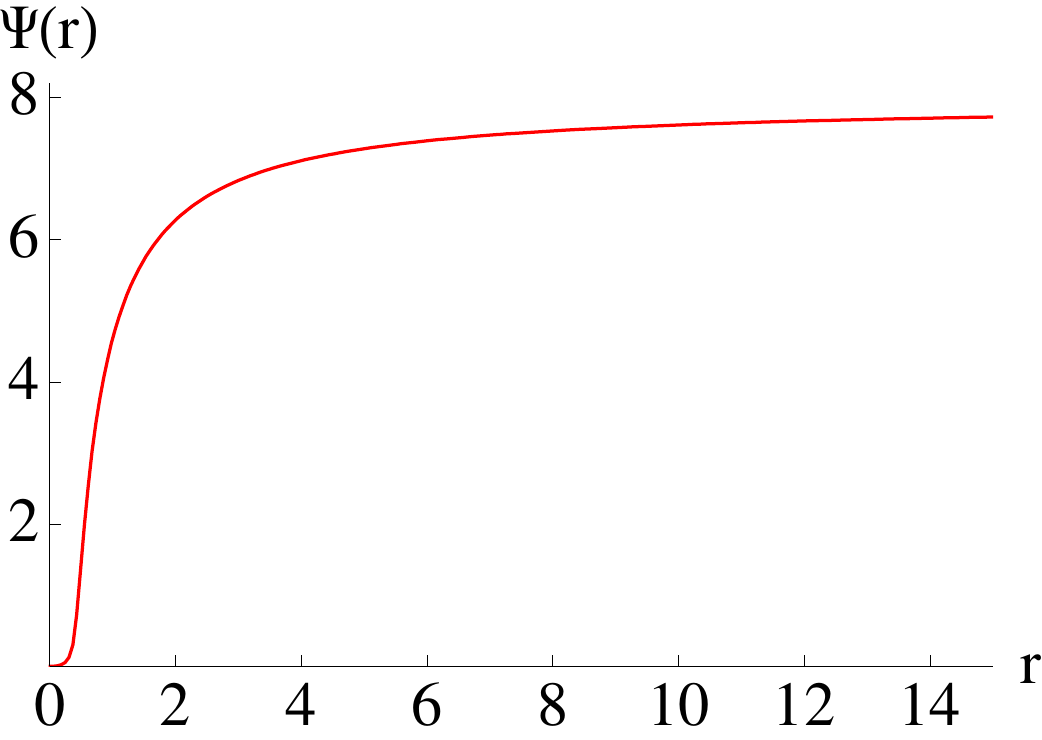}
    \caption{\small The scalar, $\lambda$, and the gauge field, $\Psi$, as functions of $r$ for the $T=0$ flow with $\lambda_1=0$ boundary condition.}
\label{O2T=0}
\end{figure}

It was argued on general grounds in \cite{Gubser:2008wz} that, at zero temperature, the solution of the Abelian Higgs model with an appropriate potential interpolates between two ${\rm AdS}_4$ spaces.\footnote{As discussed in \cite{Gubser:2009cg,Horowitz:2009ij} in general there is the possibility of having a Lifschitz solution in the IR but one can show that there are no Lifschitz solutions in the $SO(3)\times SO(3)$ truncation we study here.}  This was later realized in \cite{Gubser:2009qm,Gauntlett:2009dn,Gubser:2009gp,Gauntlett:2009bh}, where a zero temperature domain wall, which is dual to spontaneous symmetry breaking of a $U(1)$ symmetry, was constructed in consistent truncations of IIB and eleven-dimensional supergravity. It is natural to ask whether the $SO(3)\times SO(3)$ truncation discussed above admits such zero temperature solutions. One can show that such solutions indeed exist for both choices of boundary conditions, $\lambda_1=0$ or $\lambda_2=0$. To find these solutions we have again used a numerical shooting technique and have specified initial condition in the IR, which is now at $r\to0$, since there is no black hole horizon at zero temperature. To specify initial conditions in the IR we use the linearized solution of the equations of motion for $r\to 0$
\begin{align}
\begin{split}
\lambda & =  \log(2+\sqrt{5}) + \lambda^1 r^{\alpha}+\ldots \, ,\\
\Psi &= \Psi^1 r^{\beta}+\ldots \, ,\\
G &= \ds\frac{14}{3} r^2 +\ldots\,,\\
\chi &= \chi^0+\ldots\,.
\end{split}
\end{align}
where we have defined
\begin{equation}
\alpha \equiv \ds\sqrt{\ds\frac{303}{28}} - \ds\frac{3}{2}~, \qquad\qquad \beta \equiv  \ds\sqrt{\ds\frac{247}{28}} - \ds\frac{1}{2}~.
\end{equation}
Using the scaling symmetry of the equations of motion one can fix the values\footnote{We choose to work with $\Psi^1=1$ and $\chi^{0}=4$.} of $\Psi^1$ and $\chi^{0}$ and thus the only free parameter one can vary in the IR is the coefficient $\lambda^1$. We use this initial condition as a knob to set either $\lambda_1=0$ or $\lambda_2=0$ in the UV. The scalar and the gauge field for these solutions are plotted in Fig.~\ref{O1T=0} and Fig.~\ref{O2T=0}. These zero-temperature solutions should be interpreted as the ground states of the holographic superconductors with $\lambda_1=0$ or $\lambda_2=0$ boundary conditions.

%%%%%%%%%%%%%%%%%%%%%%%%%%%%%%%%%%%%%
%%%%%%%%%%%%%%%%%%%%%%%%%%%%%%%%%%%%%
%%%%%%%%%%%%%%%%%%%%%%%%%%%%%%%%%%%%%
%%%%%%%%%%%%%%%%%%%%%%%%%%%%%%%%%%%%%

%%%%%%%%%%%%%%%%%%%%%%%%%%%%%%%%%%%%%%%%%%%%%%%%%%%%%%%%%%%%%%%%%%%%%%%%%%
\section{A family of interpolating potentials}
\label{Sect:Family}
%%%%%%%%%%%%%%%%%%%%%%%%%%%%%%%%%%%%%%%%%%%%%%%%%%%%%%%%%%%%%%%%%%%%%%%%%%

We now generalize the discussion above by considering the same action as in \eqref{actionsimple}:
\begin{equation}
e^{-1}{\cal L}\eql {1\over 2}R-{1\over 4}F_{\mu\nu}F^{\mu\nu} - \partial_{\mu}\lambda\partial^{\mu}\lambda - \ds\frac{\sinh^2(2\lambda)}{4}(\partial_{\mu}\varphi -g A_{\mu})(\partial^{\mu}\varphi -g A^{\mu}) - \mathcal{P}\,,
\end{equation}
but now with a family of phenomenological potentials given by:
\begin{align}
\label{interpot1b}
\begin{split}
\mathcal{P} &=-  \ds\frac{g^2}{2} \cosh^4(\lambda) (3 - 4 \tanh^2(\lambda)) - a \ds\frac{3 g^2}{2} (2 + \cosh(2\lambda) )\\
&= - \frac{(3\,a +1)}{2} g^2 \,(3\, \cosh^4 \lambda - 4 \, \cosh^2 \lambda \, \sinh^2\lambda ) ~-~ \frac{3\,a }{2} g^2 \, \sinh^4 \lambda \,,
\end{split}
\end{align}
for some parameter, $a$.  For $a=1$ the potential reduces to  \eqref{SO3pot} in the $SO(3)\times SO(3)$ sector.   For $a=0$ the potential is the one of the $SU(4)$ sector of gauged supegravity \cite{Bobev:2010ib}, a holographic superconductor in this sector was studied in \cite{Gauntlett:2009dn,Gauntlett:2009bh}.\footnote{The action is equivalent to the one in Section 3.1 of \cite{Bobev:2010ib} with $\eta_{BHPW}=0$ and the redefinition $g_{\text{here}}=2 g_{\text{BHPW}}$.} The reason we are interested in studying this action is to illustrate how the superconducting phase transition for the condensate $O_2$ smoothly transforms from second to first order as one varies the parameter $a$. 

It is important to emphasize that the interpolating potential \eqref{interpot1b} has the following series expansion around $\lambda=0$
\begin{equation}\label{interpot2}
\mathcal{P} = - g^2(1+3a) \left( \ds\frac{3}{2} + \lambda^2 \right) + \mathcal{O}(\lambda^4)~.
\end{equation}
This implies that the scalar, $\lambda$, for all values of the parameter $a$ has the dimensionless mass $m^2L^2 = -2$ , where $L$ is the scale of the ${\rm AdS}_4$ in the UV. The only role of the parameter, $a$, at the linearized level is to determine the particular value of $L$. Therefore the parameter $a$ does not affect the linearized UV action, that is, the mass and charge of the scalar are independent of $a$.  The importance of the parameter, $a$, is that it sets the depth of the non-trivial critical point of the potential and determines the steepness of the descent to that point.

\begin{figure}[!ht]
\begin{center}
\includegraphics[width=8.5cm]{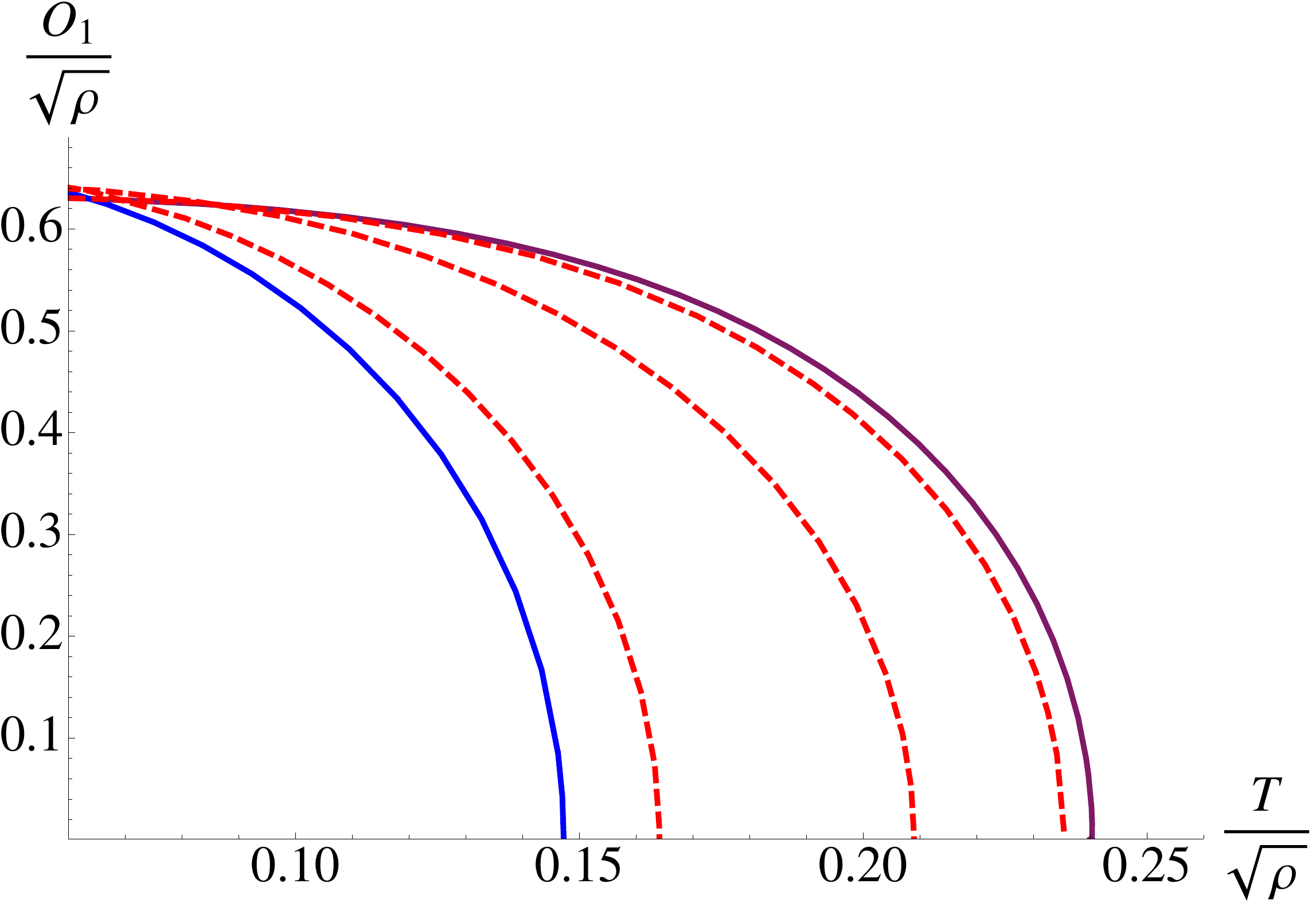}
\caption{ \small The phase diagram for the $O_1$ condensate for five representatives of the one-parameter family of interpolating solutions. The maroon (rightmost) one is for the $SO(3)\times SO(3)$ potential, that is, $a=1$. The blue (leftmost) curve is for the $SU(4)$ potential, that is, $a=0$. The three dashed curves in the middle have $a=\{0.9, 0.5, 0.1\}$  from right to left.}
\label{o1interpolation}
\end{center}
\end{figure}

\begin{figure}[!ht]
\begin{center}
\includegraphics[width=8.5cm]{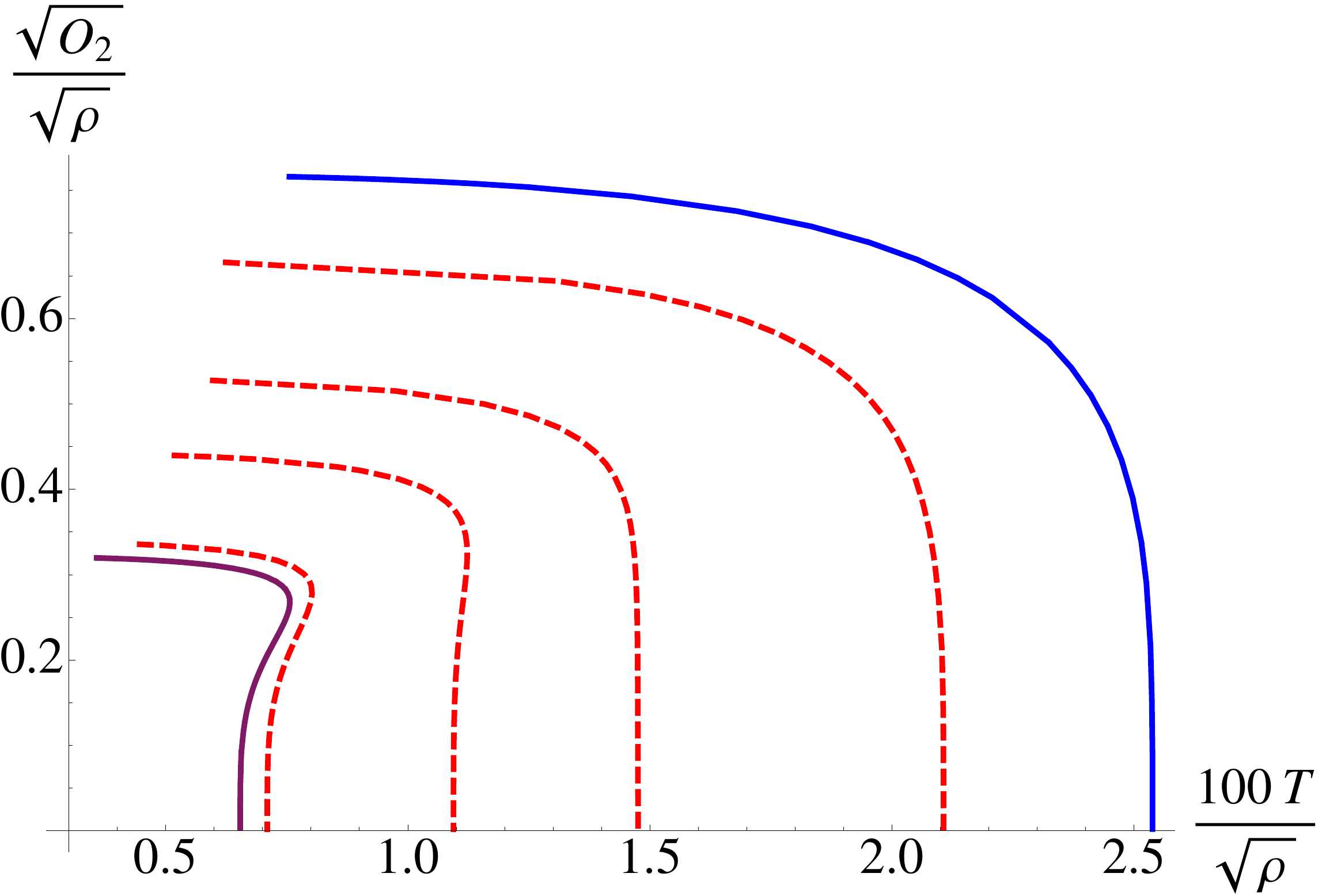}
\caption{ \small The phase diagram for the $O_2$ condensate for six representatives of the one-parameter family of interpolating solutions. The maroon (leftmost) one is for the $SO(3)\times SO(3)$ potential ($a=1$). The blue (rightmost) curve is for the $SU(4)$ potential ($a=0$). The four dashed curves in the middle have $a=\{0.9, 0.5, 0.3, 0.1\}$  from left to right. Clearly the order of the phase transition changes from first to second as $a$ is decreased.}
\label{o2interpolation}
\end{center}
\end{figure}

We have plotted the phase diagrams for $O_1$ and $O_2$ for some particular values of $a$ in Fig.~\ref{o1interpolation} and Fig.~\ref{o2interpolation} respectively. This clearly shows that the full non-linear form of the potential, which we deduced from gauged supergravity, is crucial for capturing the physics of the holographic superconductor. In particular, the phase transition for $O_2$ depicted in Fig.~\ref{o2interpolation}  changes between second order and first order as $a$ becomes larger. The phase transition for the the condensate $O_1$ remains second order for any value of $a \in [0,1]$.  One should also note that, due to lack of numerical precision in our IR shooting procedure, we are not able to find solutions with non-zero $O_2$ condensate for very low values of the temperature. We believe this is due to our imprecise numerics and is not physical. 

%%%%%%%%%%%%%%%%%%%%%%%%%%%%%%%%%%%%%
%%%%%%%%%%%%%%%%%%%%%%%%%%%%%%%%%%%%%

%%%%%%%%%%%%%%%%%%%%%%%%%%%%%%%%%%%%%
%%%%%%%%%%%%%%%%%%%%%%%%%%%%%%%%%%%%%

%%%%%%%%%%%%%%%%%%%%%%%%%%%%%%%%%%%%%
\section{The holographic dual}
%%%%%%%%%%%%%%%%%%%%%%%%%%%%%%%%%%%%%

%%%%%%%%%%%%%%%%%%%%%%%%%%%
\subsection{The standard, ``top-down'' holographic dictionary}
%%%%%%%%%%%%%%%%%%%%%%%%%%%

From the standard holographic dictionary of $\cN=8$ supergravity we know that supergravity scalars are dual to bosonic bilinears in the M2 brane theory while supergravity pseudoscalars are dual to fermion bilinears.  At the  maximally supersymmetric ${\rm AdS}_4$ critical point, all seventy supergravity scalars have masses, $m$, obeying:
\begin{equation}
m^2 \, L^2  ~=~ -2 \label{allmass}\, ,
\end{equation} 
where $L$ is the radius of curvature of the  ${\rm AdS}_4$.  This is related to the scaling dimension, $\Delta$, of the couplings or operator vevs on the M2 brane via the relation
$m^2 L^2 = \Delta(\Delta-3)$  and so the supergravity fields correspond to couplings or  vevs of operators with $\Delta =1$ or $\Delta =2$.

The standard dogma in holographic theories is that non-normalizable supergravity modes  correspond to coupling constants in perturbations of the Lagrangian of the dual theory, while normalizable supergravity modes correspond to states of the field theory on the M2 brane, described by vevs.  However, as discussed in \cite{Klebanov:1999tb}, this standard dogma does not  necessarily apply for a certain range of scalar masses and there is an ambiguity in the choice of quantization scheme of the supergravity scalar modes. The value $m^2L^2=-2$ in $\rm{AdS}_4$ falls precisely in this range. In standard quantization, the standard dogma applies but there is an ``alternative quantization''   that reverses the dictionary with non-normalizable modes describing vevs and normalizable modes representing perturbations of the Lagrangian.  Moreover, it was shown in  \cite{Breitenlohner:1982jf} that to preserve the supersymmetry in $\cN=4$ supergravity (and therefore to preserve the supersymmetry in $\cN=8$ supergravity) the supergravity pseudoscalars must be quantized in exactly the opposite way to the supergravity scalars.  Thus, if the supergravity scalars obey the standard dogma then the supergravity pseudoscalars must have the opposite dictionary, and vice versa.  \

Therefore, there are two choices of holographic dictionary for the seventy spin-$0$ particles of supergravity.  However there is only one choice in which the scaling dimensions of the supergravity modes matches precisely with the scaling dimensions of the operators or couplings of the dual M2 brane theory. The correct holographic dictionary is thus:
\begin{itemize}
\item The non-normalizable ($\Delta =1$)  modes of the $35$  pseudoscalars  describe fermion masses on the M2 brane  while for the $35$ scalars the $\Delta =1$ modes correspond to vevs of boson bilinears.

\item The normalizable ($\Delta =2$) modes of the $35$  pseudoscalars describe vevs of fermion bilinears on the M2 brane  while for the $35$ scalars the $\Delta =2$ modes correspond to bosonic masses.
\end{itemize}
This is the {\it only} dictionary that is consistent with the following three features of the maximally symmetric $\rm{AdS}_4$ vacuum (where all the supergravity scalars and pseudoscalars vanish) and the Hilbert space erected on it:  a)   $\cN=8$ supersymmetry,   b)  the relationship between supergravity scalars and  bosonic couplings/vevs on the M2 brane and  supergravity pseudoscalars and fermionic couplings/vevs on the M2 brane, and c)  the scaling dimensions of supergravity fields match the scaling dimensions of dual couplings or vevs.

To understand this holographic dictionary in more detail, one starts from the $\cN=8$ theory in the $SU(8)$ frame in which  the supersymmetries transform as the $8_s$ of $SO(8)$.  The scalars and pseudoscalars transform in the $35_v$ and $35_c$, respectively, of $SO(8)$ and can be represented by a complex, self-dual four-form, $\Sigma_{IJKL}$:
\begin{equation}
\label{selfdual}
\Sigma_{IJKL} ~=~  \coeff{1}{24}\, \epsilon_{IJKLMNPR}\, \Sigma^{MNPR}\,,
\end{equation}
where $\Sigma^{IJKL}$ is the complex conjugate of $\Sigma_{IJKL}$ and $I,J, \ldots = 1, \ldots, 8$.  The real parts of $\Sigma$ are scalars and the imaginary parts  of $\Sigma$ are the pseudoscalars.  To get the  $\cN=5$ theory one simply imposes $SO(3)$-invariance where the $SO(3)$ acts on the indices $(6,7,8)$.  The $SO(3)$-invariant scalars are thus:
\begin{equation}
\label{SO3invscal}
\phi_i \quad \leftrightarrow\quad  \Sigma_{i678}  ~=~   \Sigma^{jk\ell m}   \,,
\end{equation}
where $i=1, \dots,5$ and $(i, j, k, \ell, m)$ is an even permutation of $(1,2,3,4,5)$. Note that $\phi_1$ and  $\phi_2$ are precisely the scalars defined in (\ref{otherparam}) and that $\phi_1$ is invariant under the $SO(4) \times SO(4)$ that acts on the index sets $(2,3,4,5)$ and $(1,6,7,8)$ and  $\phi_2$ is invariant under the $SO(4) \times SO(4)$ that acts on the index sets $(1,3,4,5)$ and $(2,6,7,8)$.  If both $\phi_1$ and $\phi_2$ are non-zero then the $SO(8)$ is broken to $SO(3) \times SO(3)$ that acts on the index sets $(3,4,5)$ and $(6,7,8)$.  

One can easily use gamma matrices to convert the $35_v$ and $35_c$ representations into symmetric, traceless matrices over the  $8_v$ and $8_c$ representations.  One then finds that the real parts of  $(\phi_1,\phi_2)$ in the $35_v$  (or the imaginary parts of $(\phi_1,\phi_2)$ in the $35_c$) correspond to the matrices:
\begin{equation} \label{matform}
\left( 
\begin{matrix} a\,  \oneone_{4 \times 4}  &  b\,  \oneone_{4 \times 4}  \\  b\,  \oneone_{4 \times 4}    & - a\, \oneone_{4 \times 4} \end{matrix}  \right) \,.
\end{equation}
Note that if $a=0$ or $b=0$ then these matrices are $SO(4) \times SO(4)$-invariant but if $a, b \ne 0$ then these matrices reduce the symmetry to the diagonal $SO(4) = SO(3) \times SO(3)$. 

Setting $\zeta_1 =0$ and the phase, $\varphi=0$, implies that  $\phi_1 = \frac{1}{\sqrt{2}}  \tanh \lambda$ and $\phi_2 = \frac{i}{\sqrt{2}}  \tanh \lambda$.  One thus has one scalar and one pseudoscalar of equal magnitudes.    The operator, $\mathcal{O}_{\lambda}$,  dual to $\lambda$ is thus a linear combination of a fermion bilinear, $\mathcal{O}_{F}$,  and a boson bilinear, $\mathcal{O}_{B}$: 
\begin{equation}
\mathcal{O}_{\lambda} = \mathcal{O}_{F} +\mathcal{O}_{B} \,.
\end{equation}
We can think of the scalar $\lambda$ as a linear combination\footnote{In the full four-dimensional gauged supergravity there is of course a scalar $\widetilde{\lambda}=\lambda_B-\lambda_F$. The scalar $\widetilde{\lambda}$ is, in fact, the magnitude of the complex scalar $\zeta_1$ introduced in Section \ref{truncation}. This scalar is identically zero in our truncation.}  $\lambda=\lambda_B+\lambda_F$ where $\lambda_B$ is the scalar dual to $\mathcal{O}_B$ and $\lambda_F$ is the pseudoscalar dual to $\mathcal{O}_F$.

One can diagonalize $\mathcal{O}_{B}$ and $\mathcal{O}_{F}$ to the form:
\begin{align}
\begin{split}
\mathcal{O}_{F} &\equiv \ds\frac{1}{2\sqrt{2}} ~\text{Tr} ( \psi_1^2+\psi_2^2+ \psi_3^2+\psi_4^2 - \psi_5^2- \psi_6^2-\psi_7^2-\psi_8^2)~,\\ 
\mathcal{O}_{B} &\equiv \ds\frac{1}{2\sqrt{2}} ~\text{Tr} ( X_1^2+X_2^2+X_3^2+X_4^2 - X_5^2-X_6^2-X_7^2-X_8^2)~.
\end{split}
\end{align}
The operators $X_I$ and $\psi_A$  are, of course, the eight scalars and fermions of the $\mathcal{N}=8$ worldvolume theory on the $N$ coincident M2 branes. Both $\mathcal{O}_{F}$ and $\mathcal{O}_{B}$ are relevant operators with dimensions $\Delta_{\mathcal{O}_{F}}=2$ and $\Delta_{\mathcal{O}_{B}}=1$ and the corresponding couplings have dimensions $3 - \Delta_{\mathcal{O}_{F}}=1$  and $3 - \Delta_{\mathcal{O}_{B}}=2$.

It is also important to recall that, in five dimensions, the ``pure trace'' bilinear operators are not chiral and so do not have protected dimensions \cite{Witten:1998qj}.  This means that the holographic dictionary is ambiguous up to the addition of such operators.    Assuming that the same issue persists in four dimensions, the holographic dual of $\lambda$ is ambiguous up to the addition of the operators:
\begin{equation}
\text{Tr} \Big( \sum_{i=1}^8 \, \psi_i^2  \Big) \,,  \qquad  \text{Tr} \Big( \sum_{i=1}^8 \, X_i^2  \Big) \,. \label{ambigs}
\end{equation}

 The scalar $\lambda$   will have the following general expansion near the maximally symmetric ${\rm AdS}_4$ 
\begin{equation}
\lambda ~=~ \frac{\lambda_1}{r} ~+~ \frac{\lambda_2}{r^2}~+~ \mathcal{O}(r^{-3}) \,. 
\end{equation}
According to the holographic dictionary above, the parameter, $\lambda_1$, corresponds to a simultaneous vev of $\mathcal{O}_{B}$ and a mass insertion into the Lagrangian for  $\mathcal{O}_{F}$ and the the parameter, $\lambda_2$, corresponds to a simultaneous vev of $\mathcal{O}_{F}$ and a mass insertion into the Lagrangian for  $\mathcal{O}_{B}$.  Because of the ambiguities  (\ref{ambigs}), this could mean masses or vevs for either all eight bosons or fermions or for four of them.

A flow with boundary conditions $\lambda_1 =0$ thus corresponds to (four or eight) bosons  becoming massive and (four or eight)  fermions  developing a vev, or a condensate, while a flow with $\lambda_2=0$ describes a flow in which (four or eight) fermions are becoming massive and (four or eight) bosons are developing a vev, or condensate.   As we discussed in detail in the previous sections, the former leads to a first order phase transition while the latter leads to a second order phase transition.

%%%%%%%%%%%%%%%%%%%%%%%%%%%
\subsection{Symmetry breaking and superconductivity}
%%%%%%%%%%%%%%%%%%%%%%%%%%%

The holographic dictionary therefore implies that within the $SO(3) \times SO(3)$-invariant truncation described in Section \ref{truncation}, the flow described by $\lambda$ from the  maximally supersymmetric ${\rm AdS}_4$ critical point will  always involve a relevant perturbation of  the Lagrangian by a charged operator.  This means that the flows in $\lambda$  always involves an {\it explicit} breaking of the global $U(1)$ symmetry in the field theory on the M2 brane. Thus, even though a new condensate  subsequently develops  in the core of the solution, the $U(1)$ is explicitly and not spontaneously broken.   From the perspective of the complete $\cN=8$ theory, the mass term breaks the $SO(8)$ symmetry to $SO(4) \times SO(4)$ and the diagonal $SO(4)$ commutes with an $SO(2)$ inside $SO(8)$.  This $SO(2)$ is the gauge symmetry of our model and turning on a chemical potential for it will explicitly break the $SO(4) \times SO(4)$ to the diagonal $SO(4) = SO(3) \times SO(3)$.   Therefore, for the flows involving $\lambda$ the symmetry is fully broken at the Lagrangian level and the condensate induces no further symmetry breaking.

It should be remembered that the primary motivation for wanting spontaneous symmetry breaking is that the massless Goldstone boson will then provide the superconducting modes.  It is, of course, entirely possible that there are still superconducting modes independent of how the symmetry is broken.  Indeed,  independent of the choice of UV fixed point or quantization scheme, there is an unambiguous way to determine whether we have superconducting carriers in the dual field theory:  One can calculate the electric DC conductivity holographically using standard techniques (see for example \cite{Hartnoll:2008kx}). 

\begin{figure}[t]
 \centering
    \includegraphics[width=6.5cm]{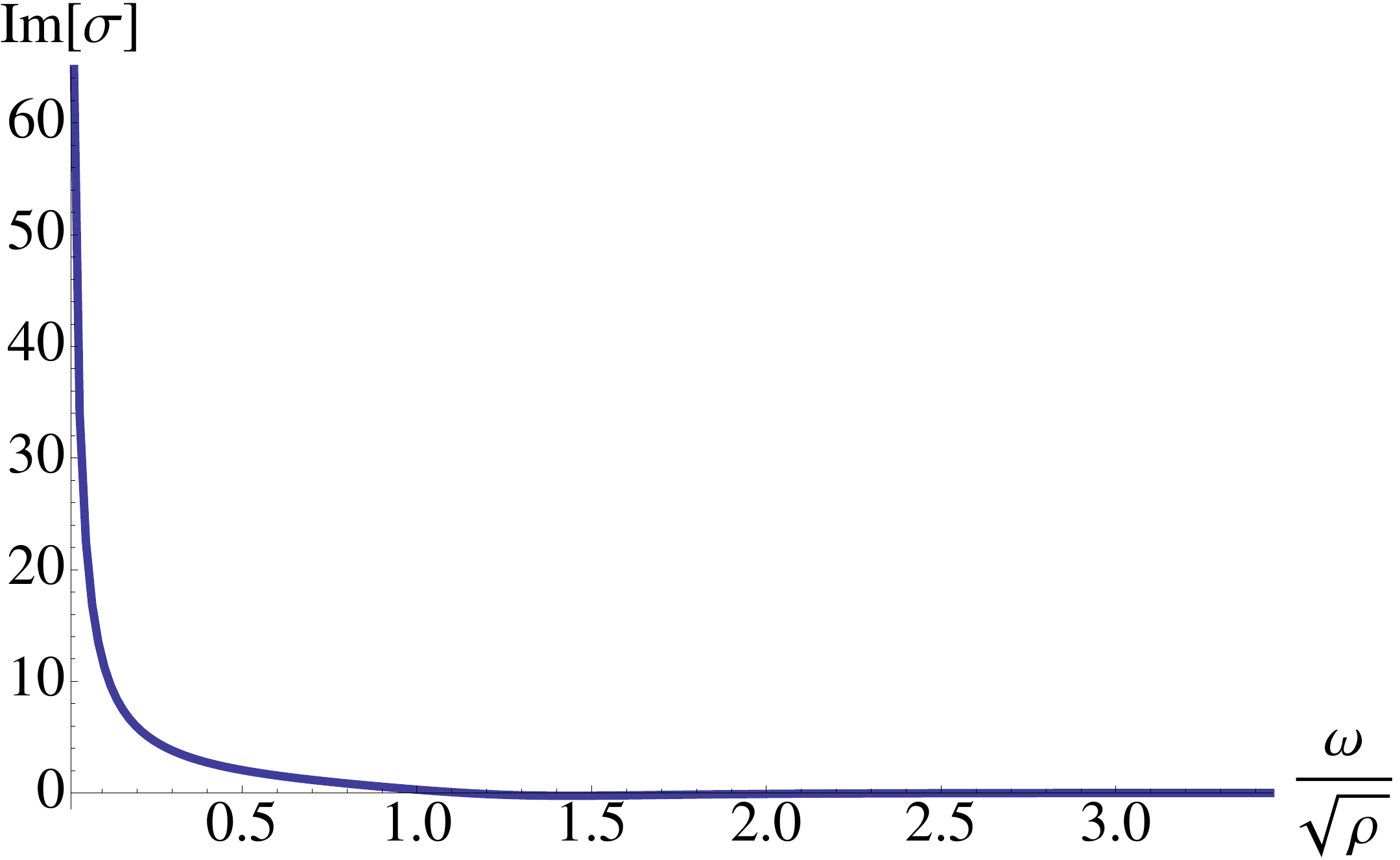}\qquad\qquad
    \includegraphics[width=6.5cm]{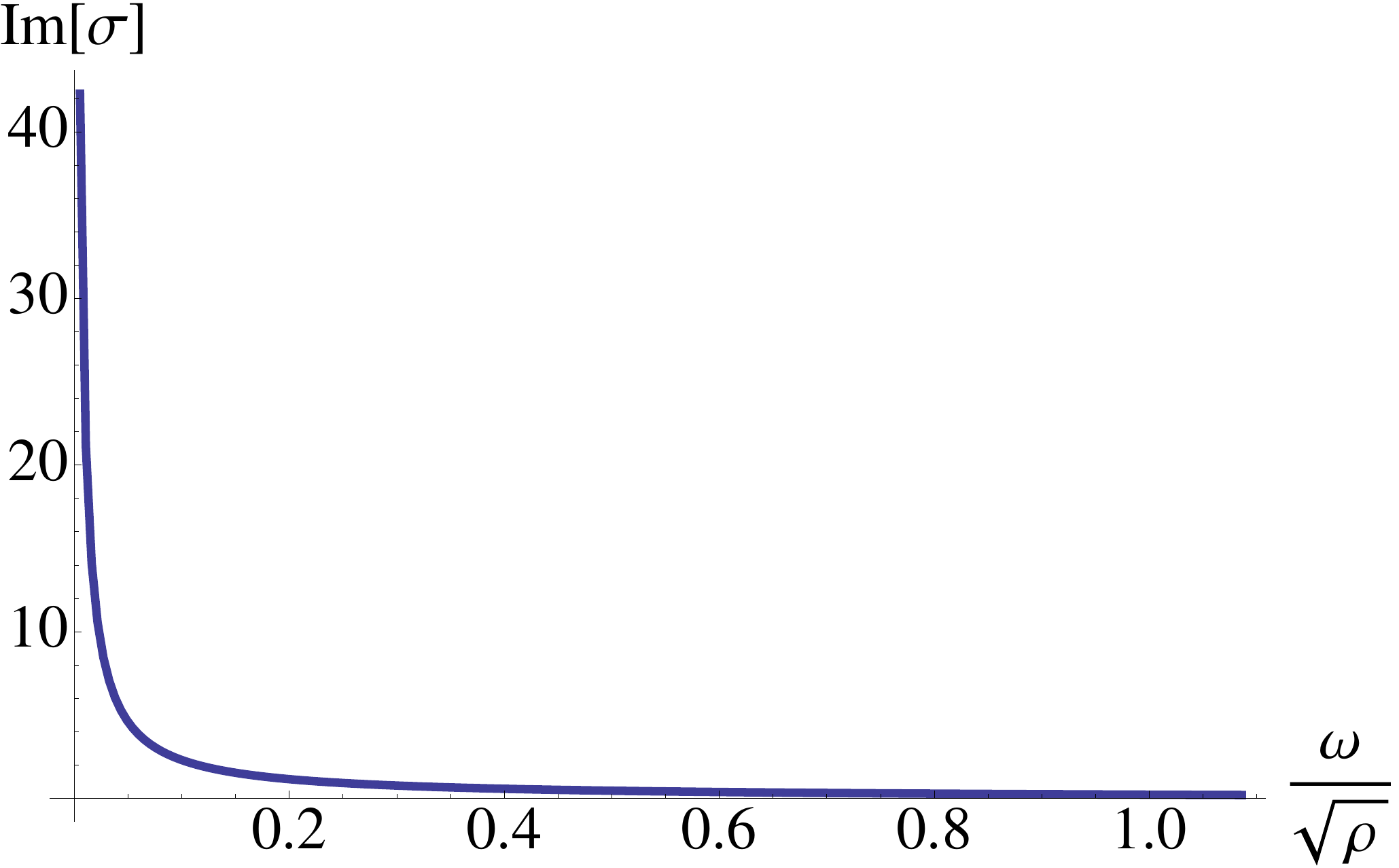}
    \caption{ \small The imaginary part of the electric conductivity as a function of the frequency for the solutions with $\lambda_2=0$ (left) and $\lambda_1=0$ (right) at $T=0$. It is clear that there is a pole at $\omega=0$.}
\label{Im[sigma]}
\end{figure}

After performing this calculation, we find that,  both for  the flows with $\lambda_1=0$ and $\lambda_2=0$, there is a delta function in the real part of the conductivity at zero frequency for temperatures less than $T_c$. This delta function cannot really be detected numerically but its presence is deduced by noticing a pole in the imaginary part of the conductivity at zero frequency and using the standard Kramers-Kronig relation. This is standard practice in similar AdS/CFT calculations and is very much along the lines of the conductivity calculations in \cite{Hartnoll:2008kx,Gubser:2008wz}.\footnote{More details on the conductivity calculation will be presented in \cite{to appear}.} To illustrate this point we have plotted the imaginary part of the electric conductivity as a function of frequency at $T=0$ in Fig.~\ref{Im[sigma]}. The behavior for other values of $T<T_c$ is qualitatively similar. The delta function in the DC conductivity clearly indicates that in the dual field theory we have superconducting (or superfluid) carriers and therefore we can unambiguously claim that our supergravity flows indeed realize holographic superconductors. This approach to detecting holographic superconductors was emphasized in \cite{Gubser:2008px}.

%%%%%%%%%%%%%%%%%%%%%%%%%%%%%%%%%%%%%
%%%%%%%%%%%%%%%%%%%%%%%%%%%%%%%%%%%%%

%%%%%%%%%%%%%%%%%%%%%%%%%%%
\subsection{An alternative UV fixed point}
%%%%%%%%%%%%%%%%%%%%%%%%%%%

There is another possible interpretation of our flows in which the dual UV fixed-point field theory is known rather implicitly but in which we should have spontaneous symmetry breaking.  One could choose to break the $\cN=8$ supersymmetry of the supergravity theory {\it ab initio}, by decorrelating the quantization of the supergravity scalars and pseudoscalars.  The resulting ``supergravity'' theory would have no supersymmetry and would not correspond to the standard $\cN=8$ superymmetric fixed-point theory on the M2 branes.  

The interpretation of this in the dual field theory is that we have deformed the theory on the M2 brane worldvolume by a relevant (or irrelevant) double-trace operator composed of the bosonic (or the fermionic) bilinear $\mathcal{O}_B$ (or $\mathcal{O}_F$). This scenario was discussed in \cite{Witten:2001ua, Berkooz:2002ug} where it was argued that these double-trace deformations will induce an RG flow to another CFT in which the scalar $\lambda_B$ (or the scalar $\lambda_F$) is in alternative quantization whereas the scalar $\lambda_F$ (or $\lambda_B$) is in standard quantization. In these new CFTs  the interpretation of the constants $\lambda_1$ and $\lambda_2$ will be different. In the CFT obtained by the double-trace deformation $\mathcal{O}^{\dagger}_B \mathcal{O}_B$, the coefficient $\lambda_2$ is dual to a sum of two vevs for the operators dual to $\lambda_B$ and $\lambda_F$, which, due to the non-trivial RG flow, are no longer the simple bosonic or fermionic bilinears of the M2 brane theory.  Similarly, in the CFT obtained by the double-trace deformation $\mathcal{O}^{\dagger}_F \mathcal{O}_F$ the coefficient $\lambda_1$ is dual to a sum of two vevs for the operators dual to $\lambda_B$ and $\lambda_F$, which again are no longer the simple bosonic or fermionic bilinears of the M2 brane theory. 

Therefore, in these new CFTs, for which we do not know the explicit Lagrangian and operator content and are therefore of limited utility, our flows with non-zero $\lambda_1$ and $\lambda_2$ will have the interpretation of spontaneous symmetry breaking flows. Following the field theory arguments in \cite{Weinberg:1986cq} these flows could be interpreted as describing a superfluid (or superconducting after ``weakly gauging" the global $U(1)$ symmetry) phase of the dual field theory.

%%%%%%%%%%%%%%%%%%%%%%%%%%%%%%%%%%%%%%%%%%%%%%%%%%%%%%%%%%%%%%%%%%%%%%%%%%
\section{Conclusions}
%%%%%%%%%%%%%%%%%%%%%%%%%%%%%%%%%%%%%%%%%%%%%%%%%%%%%%%%%%%%%%%%%%%%%%%%%%

We have found an embedding of the minimal holographic superconductor model in a consistent truncation of the maximal four-dimensional gauged supergravity. We studied finite temperature flows with non-trivial gauge field and a condensing scalar and observed that, depending on the choice of boundary condition for the scalar, $\lambda$, the superconducting phase transition could be first or second order. We also demonstrated that the zero temperature limit of our holographic superconductor is a solution that interpolates between two perturbatively stable ${\rm AdS}_4$ vacua.

We have focussed here on studying solutions of our model that exhibit the salient features of holographic superconductors. The $SO(3)\times SO(3)$ truncation is much richer and we will study it further in \cite{to appear} where we will discuss the consistent truncation of gauged supergravity in more detail and will study the uplift of the $SO(3)\times SO(3)$ critical point to eleven-dimensional supergravity. In this forthcoming work  we also find uncharged flows that realize ``triangular" RG flows (along the lines of \cite{Bobev:2009ms}) in the dual field theory and connect the three stable ${\rm AdS}_4$ vacua of the truncation in \cite{Fischbacher:2010ec}. In addition to that we find Schr\"odinger solutions of our model with irrational dynamical exponent determined completely by the value of the scalar at the $SO(3)\times SO(3)$ critical point of the scalar potential.

One of the distinguishing features of the $SO(3)\times SO(3)$ truncation is that it contains the only known\footnote{See \cite{Fischbacher:2011jx} for a recent exhaustive discussion of critical points in the $\mathcal{N}=8$ gauged supergravity.} stable, non-supersymmetric $\rm{AdS}_4$ critical point of the maximal gauged supergravity. This is important if one wants to construct minimal superconductors with well-defined zero-temperature ground states. It will be very interesting to determine whether there are other stable non-supersymmetric critical points in four dimensions and study the flow solutions in the corresponding truncation.

An important outcome of our analysis is that, for one choice of boundary conditions for the condensing scalar, we found that the superconducting phase transition is first order. This feature is due to the particular potential in the supergravity truncation we studied. We exhibited a one-parameter family of phenomenological potentials that interpolate between the $SO(3)\times SO(3)$ potential and the $SU(4)$ potential of a different embedding of the minimal holographic superconductor model in gauged supergravity \cite{Gauntlett:2009dn, Gauntlett:2009bh}. While this  family of phenomenological potentials considered in Section \ref{Sect:Family} is very interesting, it also embodies several cautionary tales for the unwary phenomenologist.  The family of solutions  for the condensate $O_2$ shows an interpolation between first order (in our model) and second order (in the model of \cite{Gauntlett:2009dn, Gauntlett:2009bh}) phase transitions. Moreover, this family involves the normalizable mode of the scalar field and so one would be very tempted to conclude, via the ``standard dogma,'' that this flow solution describes spontaneous symmetry breaking via a pure condensate with no perturbation of the Lagrangian except for a chemical potential.   At one end of the family ($a=0$) this interpretation is correct  \cite{Gauntlett:2009dn,Gauntlett:2009bh} because it can be embedded into the $\cN=8$ theory for which a precise holographic dictionary is known.  However, as pointed out in  \cite{Bobev:2010ib}  this solution is destabilized by low mass modes in supergravity and the fixed point and flow are almost certainly unphysical.  

At the other extreme ($a=1$) there is, once again, a precise holographic dictionary that also embeds the flow into the $\cN=8$ theory and this time the non-trivial fixed point has the  great virtue of being stable.  However, in spite of the normalizability of the supergravity mode, the correct holographic dictionary tells us that the flow not only involves a fermion condensate but also involves a bosonic mass term that explicitly breaks the gauge symmetry.  Therefore, while we do get a fermion condensate,  the symmetry breaking is not spontaneous but is an explicit breaking in the Lagrangian. Nevertheless we can unambiguously call this solution with a condensing scalar a holographic superconductor since it exhibits a delta function in the real part of the electric conductivity at zero frequency for $T<T_c$.

It thus seems that a minimal holographic superconductor in $2+1$ dimensions dual to $\mathcal{N}=8$ supergravity must navigate between  Scylla and Charybdis:  a non-standard, explicit symmetry breaking in the dual Lagrangian and a perturbative instability of the ground state of the system.  Whether there is a way to win through and find a ``top-down'' holographic superconductor in the $\cN=8$ theory that realizes {\it spontaneous} symmetry breaking and has a perturbatively stable ground state remains to be seen.  However, in the flow presented here we have shown, through a direct computation of the conductivity, that there is still a superconducting phase even though the symmetry breaking is not spontaneous.

%%%%%%%%%%%%%%%%%%%%%%%%%%%%%%%%%%%%%
\bigskip
\bigskip
\leftline{\bf Acknowledgements}
\smallskip
We would like to thank Jerome Gauntlett, Nick Halmagyi, Gary Horowtiz, Clifford Johnson, Igor Klebanov, Matt Roberts and Balt van Rees for useful conversations. The work of NB is supported in part by DOE grant DE-FG02-92ER-40697. The work of AK is supported by a Simons Postdoctoral Fellowship awarded by the Simons Foundation and by the National Science Foundation under Grant Number PHY-0969020. The work of KP and NW is supported in part by DOE grant DE-FG03-84ER-40168. NB and AK would like to thank the Physics and Astronomy Department at USC for hospitality while this project was initiated. NB and NW would like to thank the Aspen Center for Physics for providing a pleasant work atmosphere in the final stages of the project. NB would also like to thank the KITP-Santa Barbara for warm hospitality.

%%%%%%%%%%%%%%%%%%%%%%%%%%%%%%%%%%%%%

%%%%%%%%%%%%%%%%%%%%%%%%%%%%%%%%%%%%%

\end{document}